%% file: phylogenetic_network.tex
\documentclass[11pt]{article}
\usepackage{fullpage}
\usepackage{latexsym,amsmath,amssymb}
\usepackage{graphicx,subfigure}
\usepackage[ruled,vlined]{algorithm2e}
\usepackage{enumerate}
\usepackage{enumitem}
\usepackage{color}

\newtheorem{definition}{Definition}
\newtheorem{lemma}{Lemma}
\newtheorem{corollary}{Corollary}
\newtheorem{theorem}{Theorem}
\newtheorem{example}{Example}

\newenvironment{pf}{\begin{trivlist}
\item[\hspace{\labelsep}{\em\noindent Proof: }]
}{\hfill$\Box$\end{trivlist}}

\begin{document}

\title{Constructing a Minimum-Level Phylogenetic Network from a Dense Triplet Set in Polynomial Time}

\author{Michel Habib\thanks{Universit\'e Paris Diderot - Paris 7, LIAFA, Case 7014,
    75205 Paris Cedex 13, France.  E-mail:
  michel.habib/thu-hien.to/@liafa.jussieu.fr.}~,~~Thu-Hien To}

\maketitle

\begin{abstract}
For a given set $\mathcal{L}$ of species and a set $\mathcal{T}$ of triplets on $\mathcal{L}$, we want to construct a phylogenetic network which is consistent with $\mathcal{T}$, i.e which represents all triplets of $\mathcal{T}$. The level of a network is defined as the maximum number of hybrid vertices in its biconnected components. When $\mathcal{T}$ is dense, there exist  polynomial time algorithms to construct level-$0,1,2$ networks \cite{ASSU81,JNS06,JS06,IKKS08}. For higher levels, partial answers were obtained in \cite{IK08} with a polynomial time algorithm for simple networks. In this paper, we detail the first complete answer for the general case, solving a problem proposed in \cite{JS06} and \cite{IKKS08}: for any $k$ fixed, it is possible to construct a minimum level-$k$ network consistent with $\mathcal{T}$, if there is any, in time $O(|\mathcal{T}|^{k+1}n^{\lfloor\frac{4k}{3}\rfloor+1})$ \footnote{This is an improved result of a preliminary version  presented at CPM'2009 \cite{TH09}}.
\end{abstract}


\section{Introduction}
The goal of phylogenetics is to reconstruct plausible evolutionary histories of currently living organisms from biological data. To describe evolution, the standard model is a phylogenetic tree in which each leaf is labeled by a species, or a sequence and in which each node having descendants represents a common ancestor of its descendants. However this model is not pertinent for capturing the hybridization, recombination and lateral gene transfer events. So a new  model of network was introduced, which allows a species to have more than one parent,  see \cite{AVP}. In recent years, a lot of work has been done on developing methods for computing phylogenetic networks \cite{S2006,GHE2007,N2009,G2010}. In \cite{CJSS05}  a   parameter was introduced for phylogenetic networks, which is the number of hybridization nodes per biconnected component and called the level. The level of a network measures its distance to a tree. 

It is always difficult to reconstruct the evolution on all data set, so normally it is done on only smaller data.  Therefore, it is necessary to recombine them together into one model. A triplet is the smallest tree that contains  information on evolution, so a classic problem is to recombine a set of triplets. If there is no constraint on the triplet set, the problem  of constructing a level-$k$ phylogenetic network consistent with a triplet set is NP-hard for all levels higher than $0$ \cite{JNS06,IKKS08,IKM08}. However if the triplet set is dense, that is if we require that there is at least one triplet in the data for each three species, then the species set is better structured and then it is possible to construct a level-$1$ \cite{JNS06,JS06}, or a level-$2$ \cite{IKKS08} network, if one exists, in polynomial time. The following question was first asked  in \cite{JS06}: Does the problem remain polynomial  for level-$k$ network for a fixed $k$? We present here an affirmative answer to this question. Our preliminary version in \cite{TH09} proved that we can construct a minimum level-$k$ network in time $O(|\mathcal{T}|^{k+1}n^{3k+1})$. In this version, we present an improved result with a complexity of $O(|\mathcal{T}|^{k+1}n^{\lfloor\frac{4k}{3}\rfloor+1})$. As a consequence, it is possible to find a network with the minimum level in polynomial time if the minimum level is restricted. It means that the complexity is a polynomial function with the power of the minimum level. 
 
\subsubsection*{Related works:} \cite{ASSU81} presented an $O(|\mathcal{T}|.n)$-time algorithm for determining whether a given
set $\mathcal{T}$ of triplets on $n$ leaves is consistent with some rooted, distinctly leaf-labeled tree, i.e. a level-$0$ network, and if so, returning
such a tree. Later, improvements were given in \cite{GJLO98,HKW99}. But the problem has been proved to be  NP-hard for all other levels
\cite{JNS06,IKKS08,IKM08}. Similarly  the problem of finding a network consistent with the maximum number of triplets is also NP-hard for all levels
\cite{JNS06,IKM08}. The approximation problem which gives a factor on the number of triplets that we can construct a network consistent with, has been also
studied in \cite{BGHK08} for level-$0$, level-$1$, and level-$2$ networks.

Concerning the particular case of  dense triplet sets, there are following results. For level-$1$, \cite{JNS06} give an $O(|\mathcal{T}|)$-time algorithm
to construct a consistent network, and \cite{IK08} gives an $O(n^5)$-time algorithm to construct a consistent one with the minimum number of
reticulations. For level-$2$, \cite{IKKS08} gives an $O(|\mathcal{T}|^{\frac{8}{3}})$-time algorithm to construct a consistent network, and \cite{IK08}
presents an $O(n^9)$-time algorithm to construct the consistent one with the minimum number of reticulations. For level-$k$ networks with any fixed $k$,
there is only a result constructing all simple networks in $O(|\mathcal{T}|^{k+1})$-time \cite{IK08}. Recently, in \cite{IK11} it was proved that when the level is unrestricted, the problem is NP-hard. Besides an interesting recursive construction of level-$k$ phylogenetic networks was proposed in \cite{GBP09}.
Moreover the problem of finding a network consistent with the maximum number of triplets is NP-hard for all levels \cite{IKM08}. 

There are also studies on the version of extremely dense triplet sets, that is when $\mathcal{T}$ is considered to contain all triplets of a certain network. In this case, an $O(|\mathcal{T}|^{k+1})$ time algorithm was given in \cite{IK08} for level-$k$ networks.

\input{Preliminary}

\input{SplitSN_set}

\input{Bound}

\input{Construction}

\input{Conclusion}

\section*{Aknowledgements:}
The authors wish to thank P. Gambette for many discussions on the subject and his careful reading of this version.

\bibliography{Thesis}
\bibliographystyle{plain}

\end{document}

%% file: Preliminary.tex
\section{Preliminaries}
\label{sec:prem_triplets}
Let us  recall here some useful definitions also used in \cite{CJSS05,JNS06,JS06,IKKS08}. 
Let $\mathcal{L}$ be a set of $n$ species or taxa or sequences. 

\begin{definition}
A \textit{phylogenetic network} $N$ on $\mathcal{L}$ is a connected, directed, acyclic graph which has:

- a unique vertex of indegree $0$ and outdegree $2$ (root).

- vertices of indegree $1$ and outdegree $2$ (speciation vertices).

- vertices of indegree $2$ and outdegree $1$ (hybrid vertices or reticulation vertices).

- $n$ vertices labeled distinctly by $\mathcal{L}$ of indegree $1$ and outdegree $0$ (leaves). So $\mathcal{L}$ is also called the leaf set.
\end{definition}

\begin{figure}[ht]
\begin{center}
  \label{fig:preliminaries}
  \subfigure[A level-$2$ network $N$. \label{fig:level_2}]{\includegraphics[scale=.32]{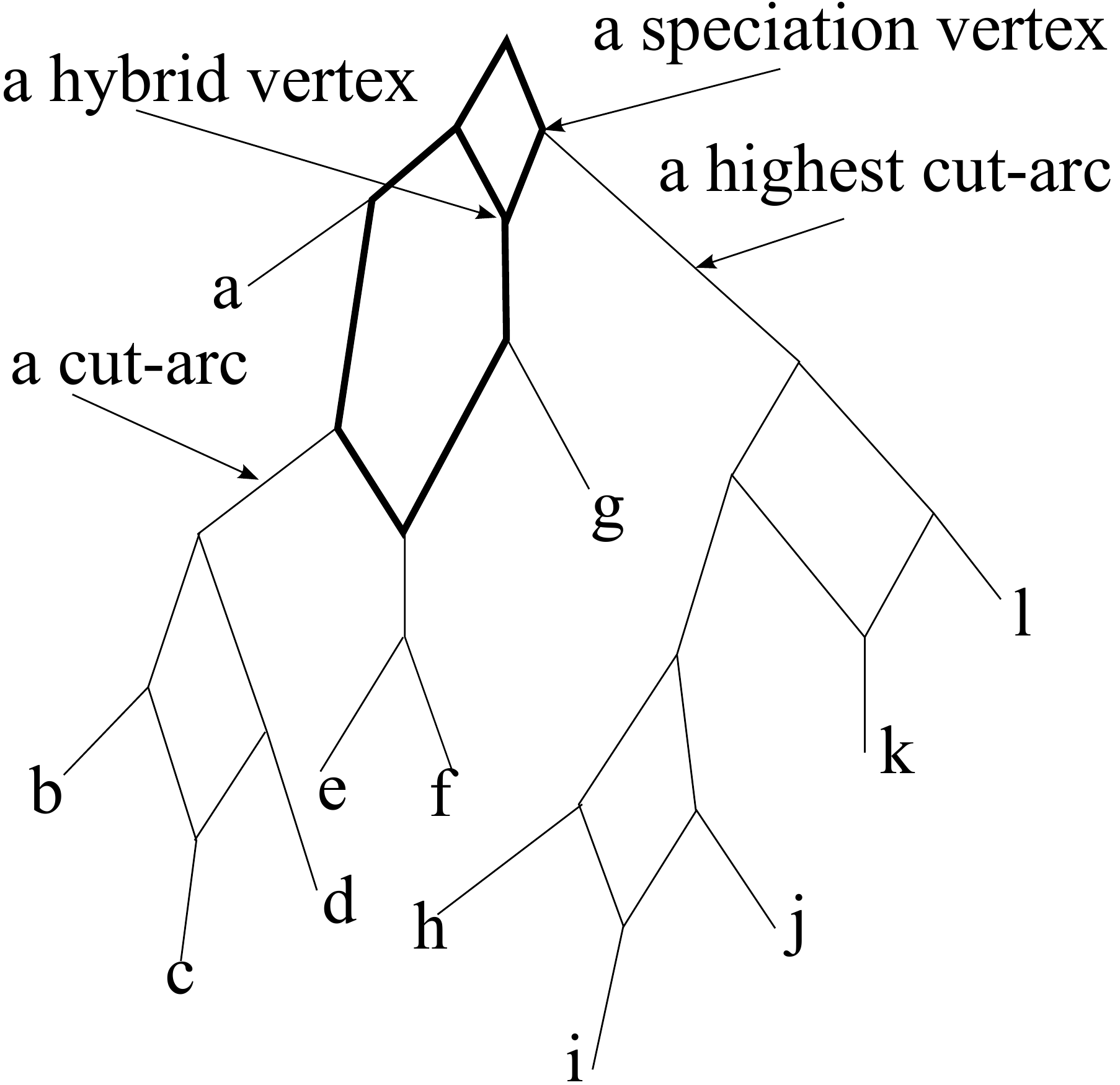}}
  \subfigure[$N_S$ \label{fig:simple_n}]{\includegraphics[scale=.32]{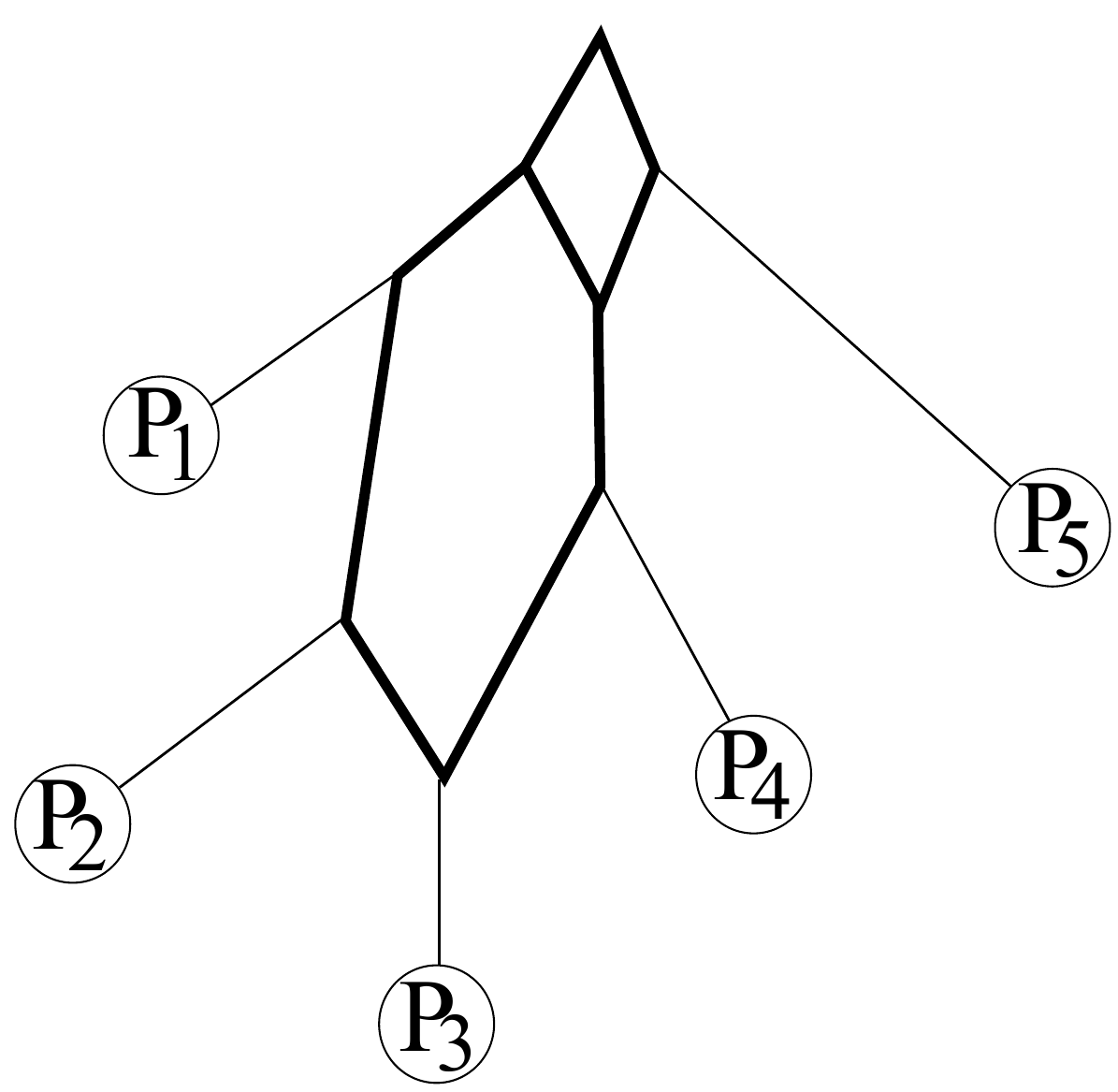}}
  \caption{A level-$2$ network and its simple network \index{simple network}.}
\end{center}
\end{figure}

For sake of simplicity, we do not show the direction of arcs in the figures. By convention, arcs are always directed away from the root. See Figure \ref{fig:level_2} for an example of a phylogenetic network on $\mathcal{L}=\{a,b,\dots,l\}$

For any two vertices $u, v$ of $N$, we denote $u \leadsto v$ \index{$\leadsto$} if $u, v$ are distinct and there is a path in $N$ from $u$ to $v$. In this case, we say that $u$ is \textit{above} \index{above} $v$ or equivalently $v$ is \textit{below} \index{below} $u$.
If $u$ is either below or above $v$ then $u,v$ are \index{comparable} comparable. Given two paths $p_1: u_1 \leadsto v_1$ and $p_2: u_2 \leadsto v_2$ such that $u_1$ is not on $p_2$, $u_2$ is not on $p_1$ and $p_1$, $p_2$ have common vertices, if $h$ is their highest common vertex, then $h$ must be a hybrid vertex. We say that $p_1,p_2$ \textbf{intersect} at $h$.

Denote by $\textbf{u} \twoheadrightarrow \textbf{v}$ \index{$\twoheadrightarrow$} for a path from $u$ to $v$ which does not contain any hybrid vertex below $u$ and above $v$, if such a path exists.


Let $\mathcal{U}(N)$ be the underlying undirected graph of $N$, obtained by replacing each directed edge of $N$ by an undirected edge. 

\begin{definition}\cite{CJSS05} A phylogenetic network $N$ is a \textbf{level-$k$ phylogenetic network} \index{level-$k$ network} iff each block of $\mathcal{U}(N)$ contains at most $k$ hybrid vertices.
 \end{definition}
The network in Figure \ref{fig:level_2} is of level-$2$. It is easy to see that $N$ is a level-$0$ phylogenetic network iff $N$ is a phylogenetic tree. 

The block of $\mathcal{U}(N)$ that contains the vertex corresponding to the root of $N$ is called the \index{highest block} highest block. By abuse, we call the subgraph of $N$ which induces this block the \textbf{highest block} \index{highest block} of $N$. In Figure \ref{fig:level_2}, the highest block is in bold. 
Denote by $\mathcal{H}$ the set of the hybrid vertices contained in the highest block of $N$, so $|\mathcal{H}| \leq k$. 

Each arc of $N$ whose removal disconnects $N$ is called a \textbf{cut-arc} \index{cut-arc}. A cut-arc $(u,v)$ is \textbf{highest} \index{highest cut-arc} if there is no cut-arc $(u',v')$ such that $v'\leadsto u$. It can be seen that a highest cut-arc always has its tail in the highest block.

A phylogenetic network is \textbf{simple} \index{simple network} if each of its highest cut-arcs connects a vertex of the highest block to a leaf. Figure \ref{fig:simple_n} represents a simple level-$2$ network. 

\begin{definition}
A \textbf{triplet} \index{triplet} $x|yz$ is a rooted binary tree whose leaves are $x$, $y$, $z$ such that $x$, and the parent of $y$ and $z$, are children of the root. A set $\mathcal{T}$ of triplets is \textbf{dense} \index{dense triplet set} if for any set $\{x,y,z\}\subseteq \mathcal{L}$, at least one triplet on these three leaves belongs to $\mathcal{T}$. 

A triplet $x|yz$ is \textbf{consistent} \index{consistent} with a network $N$ if it is 'in' this network, i.e $N$ contains two vertices $p \not= q$ and pairwise internal vertex-disjoint paths $p \leadsto x$, $ p\leadsto q$, $q\leadsto y$, and $q\leadsto z$.
\end{definition}

If a triplet is consistent with a network, we says also that this network is consistent with the triplet. For example, $a|bc$ is consistent with the network in Figure \ref{fig:level_2}, but $b|ac$ is not. 

A network is \textit{consistent} with a set of triplets iff it is consistent with all triplets in this set.

For the sake of simplicity, in the following it is always assumed that  $\mathcal{T}$ is a dense triplet set, and we will consider the following problem :

\vspace{0.5cm}

\textbf{Main Problem}

\textbf{data:} A dense  triplet set $\mathcal{T}$ and a fixed integer $k$.

\textbf{research:} A  level-$k$ phylogenetic network consistent with $\mathcal{T}$.

\vspace{0.5cm}

We call a level-$k$ network consistent with $\mathcal{T}$ having the minimum number of hybrid vertices a \textbf{minimum} level-$k$ network \index{minimum network} consistent with $\mathcal{T}$.

Let $L$ be a subset of the $\mathcal{L}$. The restriction of  $\mathcal{T}$ to $L$ is denoted by $\mathcal{T}|L=\{x|yz \in \mathcal{T}$ such that $x,y,z \in L\}$. 
Let $\mathcal{P}$ be a partition of $\mathcal{L}$: $\mathcal{P}=\{P_1, \dots, P_m\}$. Denote $\mathcal{T} \nabla \mathcal{P}=\{P_i|P_jP_k$ such that $\exists~x \in P_i, y \in P_j, z \in P_k$ with $x|yz \in \mathcal{T}$ and $i, j$ and $k$ are distinct$\}$ \index{$\mathcal{T} \nabla \mathcal{P}$}.

For each network $N$, by removing the highest block and the highest cut-arcs, we obtain several vertex-disjoint subnetworks $N_1, \dots , N_m$. Each one is hung below a highest cut-arc. If in $N$, we replace each $N_i$ by a leaf, then we have a simple network called $\textbf{N}_\textbf{S}$ (Figure \ref{fig:simple_n}). Let $l(N_i)$ be the leaf set of $N_i$, so a $l(N_i)$  is called a \textit{leaf set below a highest cut-arc}. It is easy to see that $\textbf{P(N)} = \{l(N_1), \dots , l(N_m)\}$ \index{$P(N)$} is a partition of $\mathcal{L}$.  We can use biconnectivity to decompose our problem as described in \cite{IKKS08}. 

\begin{lemma}\label{lem:decomposition} \textbf{Decomposition lemma}
$N$ is a level-$k$ network consistent with $\mathcal{T}$ iff each $N_i$ is a level-$k$ network consistent with $\mathcal{T}|l(N_i)$ for any $i = 1, \dots, m$ and $N_S$ is a simple level-$k$ network consistent with $\mathcal{T} \nabla P(N)$.
\end{lemma}

Constructing a simple level-$k$ network consistent with $\mathcal{T} \nabla P(N)$, if such a one exists, can be done in polynomial time using \cite{IK08}. Therefore the main difficulty if we want to derive from this lemma a divide and conquer approach is to estimate the number of partitions that have to be checked. Fortunately the search  can be restricted to a polynomial number of partitions and to this aim further definitions and technical lemmas are developed in the next sections.

\subsection{SN-sets} 

Remark that if $A$ is a leaf set below a cut-arc, i.e. a part of $P(N)$, then for any $z \in \mathcal{L} \backslash A$, $x, y\in A$, the only triplet on $\{x,y,z\}$ that can be consistent with the network is $z|xy$. Based on this remark, we define a family of leaf sets, called \textbf{CA-sets}, for CutArc-sets, as follows.

\begin{definition} \index{CA-set}
\label{def:CA_set}
Let $A \subseteq \mathcal{L}$, then $A$ is a CA-set \index{CA-set} if either it is a singleton or the whole $\mathcal{L}$, or if it satisfies the following property: For any $z \in \mathcal{L} \backslash A$, $x, y\in A$, the only triplet on $\{x,y,z\}$ in $\mathcal{T}$, if there is any, is $z|xy$.
\end{definition}

As noticed above, each part of $P(N)$ is a CA-set, but the converse claim is not always true. Let us recall  that \cite{JS06} introduced  a variation  of these CA-sets by a closure operation, namely the notion of SN-set. In \cite{TH09}, we showed the equivalence between these two definitions. 
Therefore, the family of SN-sets is exactly the family of CA-sets and  we will stick to  the notation of SN-set for any CA-set determined by Definition \ref{def:CA_set}.

It was proved in \cite{JS06} that if $\mathcal{T}$ is dense, then the collection of its SN-sets is a laminar family \index{laminar family} \cite{S03}.
It means that $2$ SN-sets are either disjoint or one of them contains the other, hence this family can be represented by a tree when considering inclusion. This tree is called \textbf{SN-tree} \index{SN-tree}, its root corresponds to $\mathcal{L}$, and the leaves correspond to the singletons. Moreover  each node $v$ of the SN-tree represents an SN-set made up with the leaves of the subtree rooted in $v$. 
Let $s_1, s_2$ be two SN-sets.
We say that $s_1$ is a \textbf{child} of $s_2$, or $s_2$ is a \textbf{parent} of $s_1$, if in the SN-tree, the node representing $s_1$ is a child of the node representing $s_2$. For example, in the SN-tree of Figure \ref{fig:split_ex}, the SN-set $\{d,e\}$ is a child of the SN-set $\{a,b,c,d,e\}$. A non trivial maximal SN-set is a child of $\mathcal{L}$. To simplify the notation, we call such a set a \textbf{maximal SN-set} \index{maximal SN-set}.

Take for example the SN-tree in Figure \ref{fig:split_ex}. The SN-set $\{f,k,h,g,j,l\}$ has two children $\{f,k\}$ and $\{h,g,j,l\}$. There are two maximal SN-sets which are $\{a,b,c,d,e\}$ and $\{i,f,k,h,g,j,l\}$.


\subsection{Split SN-sets}

\begin{definition}
Let $N$ be a network on $\mathcal{L}$ consistent with $\mathcal{T}$, and let $S$ be an SN-set of $\mathcal{T}$ different from $\mathcal{L}$. We say that $S$ is \textbf{split} \index{split SN-set} in $N$ iff each child of $S$ is equal to a part of $P(N)$. In other words, each child of $S$ is the leaf set below a highest cut-arc of $N$, or a certain $l(N_i)$.
\end{definition}

\begin{example}
\label{ex:split_SN_set}
For example, suppose that $\mathcal{T}$ has the SN-tree in Figure \ref{fig:split_ex}. $\mathcal{T}$ can be the set of all possible triplets on the leaf set $\mathcal{L} = \{a,b, \dots k,l\}$ except $x|yz$ such that $x,y$ but not $z$ are in an SN-set. It can be verified that the network $N$ in Figure \ref{fig:split} is consistent with $\mathcal{T}$. $N$ has $9$ subnetworks $N_i$ which gives us the partition $P(N)=$ $\{\{a,b,c\}$, $\{d\}$, $\{e\}$, $\{f\}$, $\{g\}$,$\{h\}$, $\{i\}$, $\{k\}$, $\{j,l\}\}$. The SN-set $\{g,h,j,l\}$ is split in this network because each of its children, $\{g\},\{h\},\{j,l\}$, corresponds to a part of $P(N)$. The SN-sets $\{d,e\}$, $\{f,k\}$ are also split \index{split SN-set} here. However, the SN-set $\{a,b,c,d,e\}$ is not split in this network. Indeed, its children are $\{a,b,c\}$, $\{d,e\}$  and the latter one is not equal to any part of $P(N)$. All other SN-sets, $\{a,b,c\}$,$\{j,l\}$,$\{f,g,h,j,k,l\}$, $\{f,g,h,i,j,k,l\}$, the singletons, and the whole $\mathcal{L}$, are neither split. In Figure \ref{fig:split_ex}, each white round node corresponds to an SN-set which is split in $N$, each square node corresponds to an SN-set which is a part of $P(N)$.

\begin{figure}[ht]
\begin{center}
  \subfigure[The SN-tree of $\mathcal{T}$. The white black square nodes represent the SN-sets that split in $N$. The white square nodes represent the partition of $\mathcal{L}$ in $N$. \label{fig:split_ex}]{\includegraphics[scale=0.5]{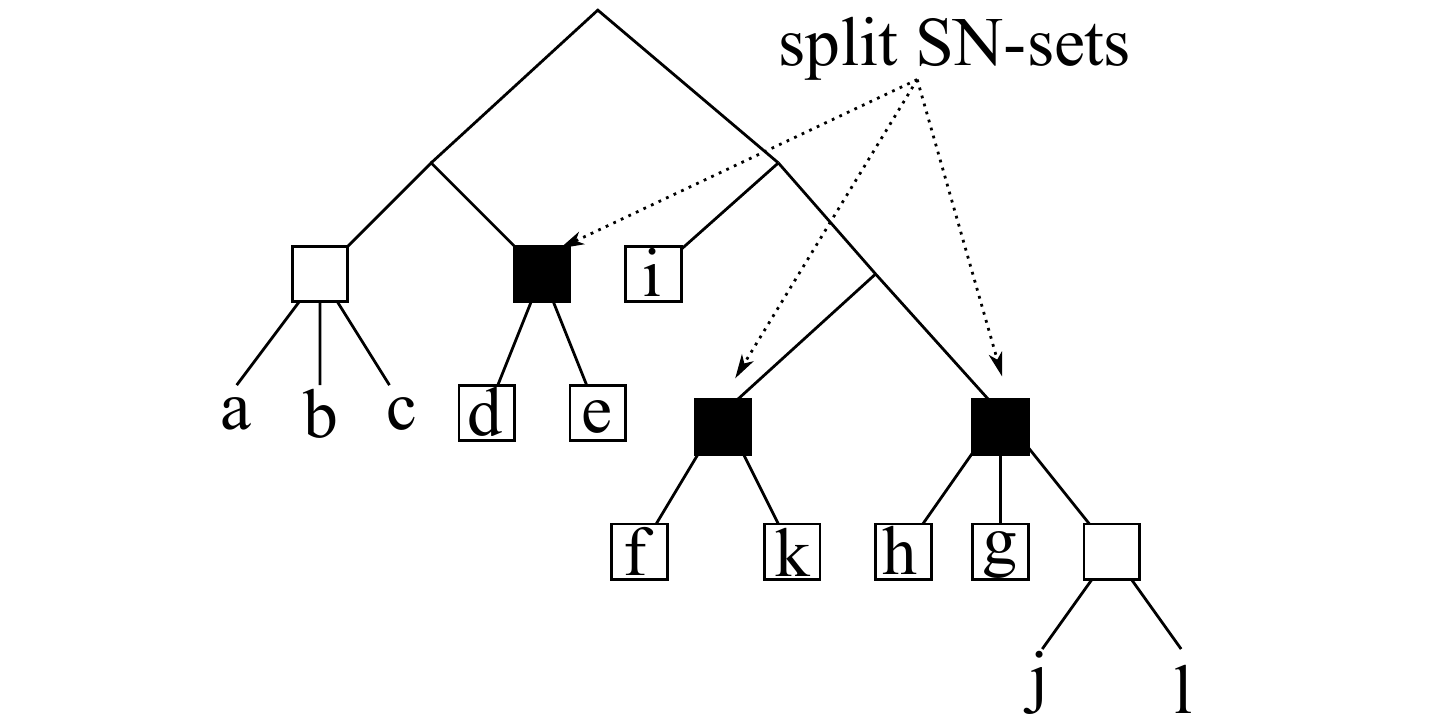}}\;\;\;\;\;
  \subfigure[A network $N$ consistent with $\mathcal{T}$.  \label{fig:split}]{\includegraphics[scale=0.5]{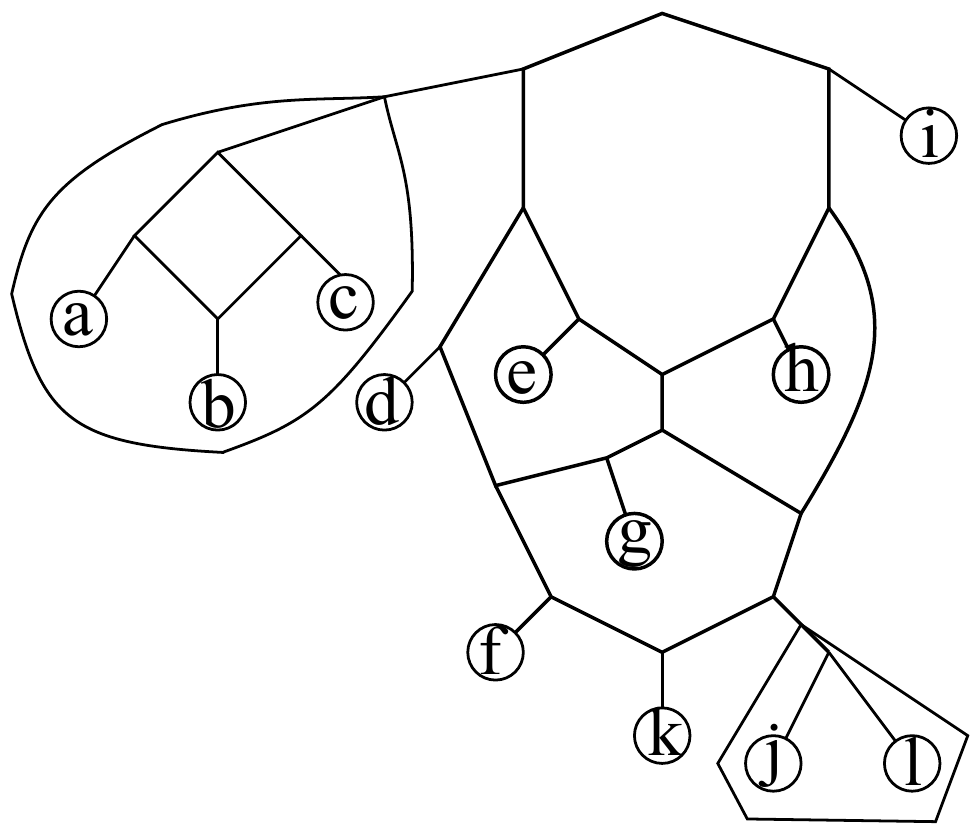}}
  \caption{Example \ref{ex:split_SN_set} of split SN-set. \index{split SN-set}}
  \label{fig:ex_split}
\end{center}
\end{figure}
\end{example}

By definition, a tree consistent with $\mathcal{T}$ has no split SN-set. 

\cite{JS06,JNS06} showed that if $\mathcal{T}$ is consistent with a level-$1$ network, then there exists a level-$1$ network $N$ consistent with $\mathcal{T}$ such that each maximal SN-set is a part of $P(N)$. So $N$ has no split SN-set.

\cite{IKKS08} showed that if $\mathcal{T}$ is consistent with a level-$2$ network, then there exists a level-$2$ network $N$ consistent with $\mathcal{T}$ in which each  maximal SN-set is a part of $P(N)$, except at most one maximal SN-set $S$ such that each child of $S$ equals a part of $P(N)$. So, $N$ has at most $1$ split SN-set.

For level-$k$ networks, with $k \ge 3$,  each part of $P(N)$ does not always correspond to a maximal SN-set and it can be any SN-set at any level in the SN-tree (its depth in the SN-tree is not bounded by a function of $k$), but the number of split SN-sets is bounded by a linear function of $k$. Indeed, in \cite{TH09}, we proved that a level-$k$ network consistent with $\mathcal{T}$ has at most $3k$ split SN-sets. In this paper, we propose a stricter bound: if $\mathcal{T}$ is consistent with a level-$k$ network, then there is a level-$k$ network $N$ consistent with $\mathcal{T}$ such that the number of split SN-sets of $N$ is bounded by $\lfloor \frac{4}{3}k \rfloor$.

It is easy to see that two SN-sets which are both split in $N$ are disjoint. 
An SN-set may be split in a network but not-split in another network which is also consistent with the same triplet set. Therefore, when we say that $S$ is split, we have to indicate in which network. However, for convenience, from now on, when we say $S$ is a split SN-set, it means that $S$ is split in $N$, a level-$k$ network consistent with $\mathcal{T}$ that we are going to construct.

\begin{lemma}
\label{lem:calculation}
The set of split SN-sets of $N$ totally determines $P(N)$.
\end{lemma}

\begin{pf}
Suppose that we know the set of all split SN-sets of a network $N$, we can determine $P(N)$ as follows.
Let $P_i$ be a part of $P(N)$. So either $P_i$ is a child of a split SN-set or not included in any split SN-set. In the latter case, it is a biggest one that is not comparable (neither included nor containing) with any split SN-set. For example, see Figure \ref{fig:split_ex} where each split SN-set corresponds to a white round node and each part of $P(N)$ corresponds to a black square node.
\end{pf}

So, to bound the number of possible partitions of consistent level-$k$ networks, we will find a bound for the number of split SN-sets in a consistent level-$k$ network. The idea is to find relations between the number of split SN-sets and the number of the hybrid vertices. To this aim, some functions from a split SN-set to a set of hybrid vertices will be introduced.

%% file: SplitSN_set.tex
\section{Some properties and functions of split SN-sets}
\label{sec:split}

This section explores some properties and functions of split SN-sets which will be used in Section \ref{sec:strict_bound} to find a stricter bound for the number of split SN-sets in a level-$k$ network consistent with $\mathcal{T}$. 

A vertex of $N$ is a lowest common ancestor, \index{lowest common ancestor} $lca$ \index{$lca$} for abbreviation, of a split SN-set $S$ if it is the $lca$ of all leaves of $S$. If $S$ has only one $lca$, then we denote it by $\textbf{lca(S)}$ \index{$lca(S)$}. Remark that a $lca$ of $S$ is never a hybrid vertex.

Let $t$ be a $lca$ of $S$, denote by $\textbf{N}_\textbf{t}\textbf{[S]}$ the induced subgraph of $N$ consisting of all paths from $t$ to the leaves of $S$, and $\textbf{N[S]} = \cup~N_t[S]$ for all $lca$s $t$ of $S$. In the figures which describe $N$ in the following,  we represent $N[S]$ by continuous lines and the parts not in $N[S]$ by dotted lines.

For any SN-set $S$ of $\mathcal{T}$ which is split in $N$, denote by $s_1, \dots , s_m$ the children of $S$, i.e. each $s_i$ is a part of $P(N)$. Let $\textbf{u}_\textbf{i}$ \index{$u_i$}  be the tail of the highest cut-arc below which  $s_i$ is attached to, so $u_i$ is on the highest block of $N$. Sometimes, we denote $u_{s_i}$ instead of $u_i$ when there are more than one split SN-sets that are involved.
For any subset $f$ of $\mathcal{L} \setminus S$, and children $s_i,s_j$ of $S$, denote $f|s_is_j= \{x|yz \in \mathcal{T}$ such that $x \in f$ and $y,z \in s_i \cup s_j\}$.

\subsection{Function $a$}

\begin{minipage}[b]{11cm}
\begin{definition}\index{function $a$}
\label{def:h}
For any split SN-set $S$ of $N$, let:

$a(S)=\{h \in \mathcal{H}~|~h$ is in $N[S]$ and $\exists$ a child $s_i$ of $S$ such that either $h=u_i$ or there is a path $h \twoheadrightarrow u_i\}$ \index{$\twoheadrightarrow$} ($a$ for above)
\end{definition}
\begin{example}
For example, in the figure on the right, $a(S) = \{h_0,h_2,u_4\}$ because they are in $\mathcal{H}$, in $N[S]$, and we have the paths $h_0 \twoheadrightarrow u_3$, $h_2 \twoheadrightarrow u_2$. $h_1$ is in $N[S]$ but it is not in $a(S)$ because any path from $h_1$ to any $u_i$ contains at least another hybrid vertex.
\end{example}
\end{minipage}\hfill
\begin{minipage}[b]{4cm}
\def\svgwidth{4cm}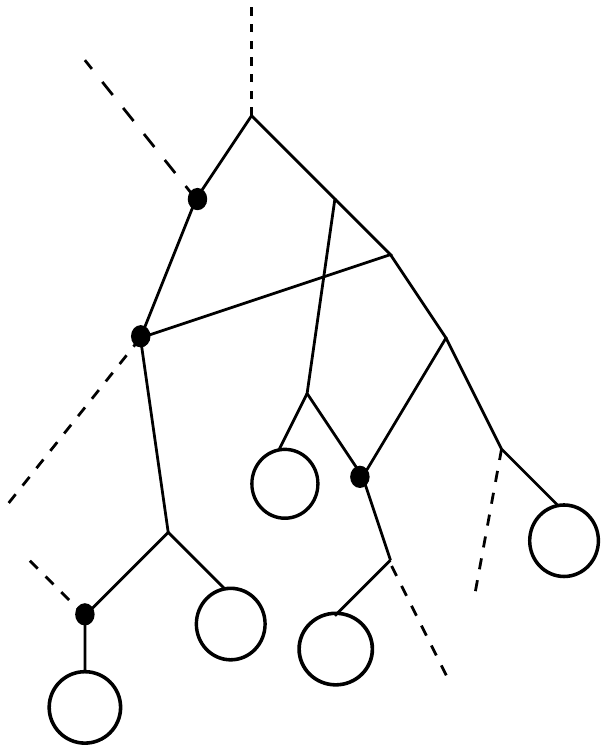
\end{minipage}

\begin{lemma}Let $S$ be a split SN-set of $N$.
\label{lem:a1}

(i) $a(S) = \emptyset$ iff $N[S]$ does not contain any hybrid vertex of $\mathcal{H}$. 

(ii) If $|a(S)| \le 1$ then $S$ has only one lca.

(iii) If $a(S) = \emptyset$, then $\forall x \not\in S$, there must exist a path from the root to $x$ which is vertex-disjoint with $N[S]$. 
\end{lemma}

\begin{pf}
(i) This claim is inferred directly from the definition of $a$. 

(ii) Suppose that $S$ has $2$ lcas, called $t_1, t_2$.
Let $s_1$ be a child of $S$, so there exists a path $t_1 \leadsto u_1$ and a path $t_2 \leadsto u_1$. 

\begin{minipage}[b]{12cm}
Since $t_1$ is neither above nor below $t_2$, these two paths must intersect at a hybrid vertex above $u_1$.
Let $h_1$ be a lowest hybrid vertex below $t_1,t_2$ and above $u_1$, i.e. we have a path $h_1 \twoheadrightarrow u_1$. So $h_1 \in a(S)$ by definition. Let $s_2$ be another child of $S$ such that $t_1, t_2$ are lcas of $s_1,s_2$. By the same argument, there is a hybrid vertex $h_2$ and a path $h_2  \twoheadrightarrow u_2$, i.e. $h_2 \in a(S)$. It is evident that $h_1 \neq h_2$ because otherwise $h_1$ is a lca of $s_1,s_2$. So $a(S)$ contains at least two hybrid vertices $h_1,h_2$, a contradiction.
\end{minipage}\hfill
\begin{minipage}[b]{3cm}
\def\svgwidth{3cm}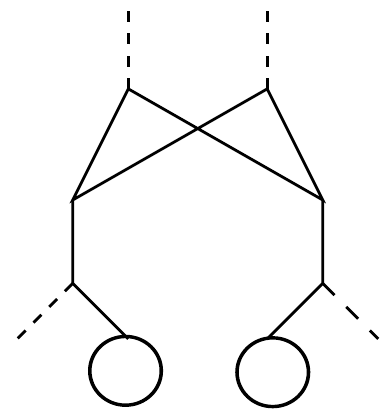
\end{minipage}

(iii) From the fact that $a(S) = \emptyset$, we deduce that  $N[S]$ does not have any hybrid vertex in $\mathcal{H}$ and $S$ has only one lca.  Let $s_i,s_j$ be two children of $S$ such that $lca(s_i,s_j) = lca(S)$.
Because $S$ is an SN-set, $x|s_is_j$ is consistent with $N$ for any $x \not\in S$.
So, there exist two vertices $p,q$ of $N$ such that there are $4$ internal vertex-disjoint paths $p \leadsto x$, $p \leadsto q$, $q \leadsto u_i$ and $q \leadsto u_j$.

\begin{minipage}[b]{11cm}
We deduce that $q=lca(S)$ because $u_i,u_j$ have only one lca which is $lca(S)$.
Suppose that the path $p \leadsto x$ has common vertices with $N[S]$, then this path must pass $lca(S)$ because $N[S]$ does not have any hybrid vertex in 
$\mathcal{H}$. It implies that the paths $p \leadsto x$, $q \leadsto u_i$, $q \leadsto u_j$ have $lca(S)$ as a common vertex, a contradiction. So, there exist at least a path $p \leadsto x$ which is vertex-disjoint with $N[S]$.
\end{minipage}\hfill
\begin{minipage}[b]{3.2cm}
\def\svgwidth{3.2cm}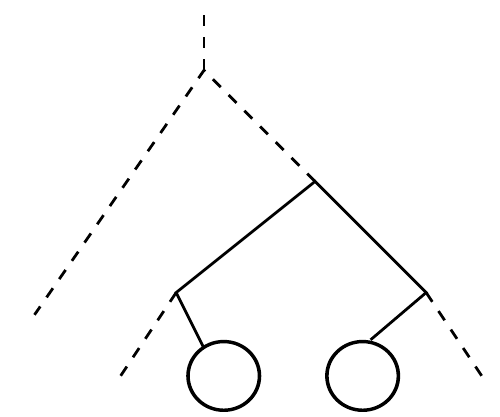
\end{minipage}
\end{pf}

\begin{lemma}
\label{lem:function}
For any $h \in \mathcal{H}$, $|a^{-1}(h)| \le 1$, i.e. a hybrid vertex is assigned to at most one split SN-set by function $a$.
\end{lemma}

\begin{pf}
Assume that there exist $2$ split SN-sets $X, Y$ such that $h \in a(X) \cap a(Y)$.

\begin{minipage}[b]{11cm}
By definition, there exists $x$ which is a child of $X$ such that either $h=u_x$ or there is a path $h \twoheadrightarrow u_x$. Similarly we have a child $y$ of $Y$. Let $t_Y$ be a lca of $Y$, so $t_Y$ is above $h$ and let $y'$ be another child of $Y$ such that  $lca(y,y') = t_Y$. We see that any paths from a vertex above $t_Y$ to $x$ and to $y$ must pass $h$ because there is no hybrid vertex on the paths from $h$ to $u_x$ and to $u_y$. So $x|yy'$ is not consistent with the network, contradicting  $Y$ being an SN-set.
\end{minipage}\hfill
\begin{minipage}[b]{4cm}
\includegraphics[scale=.6]{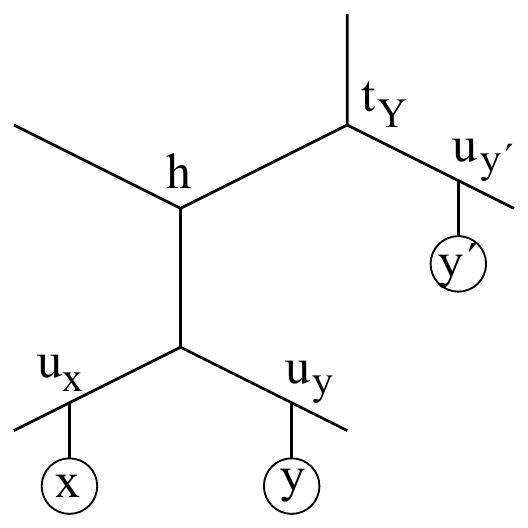}
\end{minipage}
\end{pf}

\subsection{Function $b$}

Let $S$ be a split SN-set of $N$ such that $a(S) = \emptyset$, and $h$ be a hybrid vertex not in $N[S]$ below $lca(S)$. 
We denote $\textbf{lca(S)} \hookrightarrow \textbf{h}$ \index{$\hookrightarrow$} for a path from $lca(S)$ to $h$ such that 
\textit{there is a path from the root to $h$ which is vertex-disjoint with $N[S]$, and for any hybrid vertex $h'$ above $h$ on this path, every path from the root to $h'$ has common vertices with $N[S]$}.
In other words, $h$ is a highest hybrid vertex below $lca(S)$ which has a path coming to it from outside of $N[S]$.
 
Note that if there is a path $lca(S) \twoheadrightarrow h$, i.e. if there is no hybrid vertex different from $h$ on this path, then this path is also a path $lca(S) \hookrightarrow h$.

\begin{minipage}[b]{10cm}
\begin{definition}\index{function $b$}
For any split SN-set $S$ of $N$ such that $a(S) = \emptyset$, let $b(S) = \{h \in \mathcal{H}~|~\exists$ a path $lca(S) \hookrightarrow h\}$
\index{$\hookrightarrow$}
($b$ for below)
\end{definition}

\begin{example}
For example in the figure on the right, $b(S) = \{h_1,h_2\}$ because they are in $\mathcal{H}$ and we have two paths $lca(S) \hookrightarrow h_1$, $lca(S) \hookrightarrow h_2$. $h_0$ is not in $b(S)$ because every path from the root to $h_0$ has common vertices with $N[S]$. $h_3$ is not in $b(S)$ because there is only one path from $lca(S)$ to $h_3$ but this path contains the hybrid vertex $h_2$ above $h_3$ and there is a path from the root to $h_2$ (the dotted line) which is vertex-disjoint with $N[S]$. 
\end{example}
\end{minipage}\hfill
\begin{minipage}[b]{5cm}
\def\svgwidth{5cm}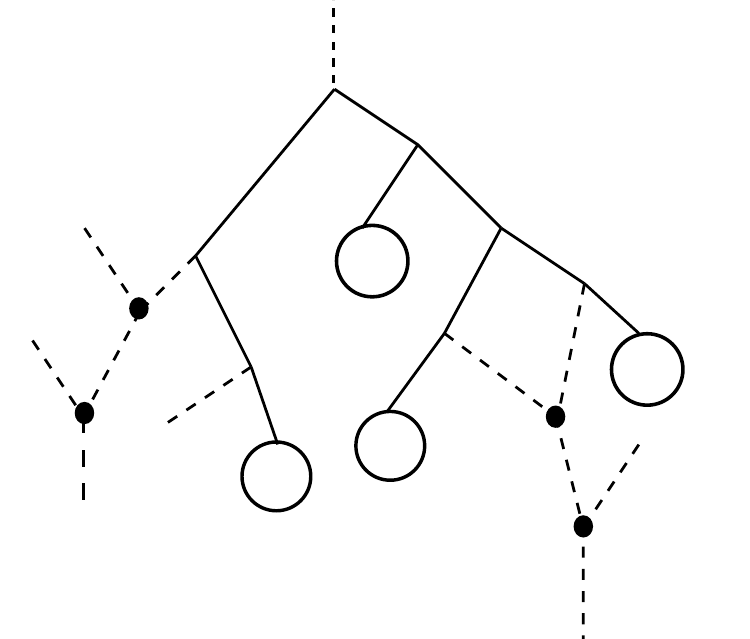
\end{minipage}

\begin{lemma}
\label{lem:function_b}
$\forall h \in \mathcal{H}$, there are at most $2$ split SN-sets $X, Y$ such that $a(X) = a(Y) = \emptyset$ and $h \in b(X) \cap b(Y)$.
\end{lemma}

\begin{pf}
Suppose that there are $3$ split SN-sets $X, Y, Z$ such that $a(X) = a(Y) = a(Z) = \emptyset$ and a hybrid vertex $h \in b(X) \cap b(Y) \cap b(Z)$. By definition, $h$ is below $lca(X)$, $h$ is not in $N[X]$ and there is a path $c_X$: $lca(X) \hookrightarrow h$. Similarly for $Y, Z$.


The $3$ paths $c_X,c_Y,c_Z$ pass $h$, then there are at least two among them, for example $c_Y, c_Z$ have common vertex above $h$. We have the following cases:

\begin{minipage}[b]{10cm}
(i) $c_Y$ and $c_Z$ intersect at a hybrid vertex $h'$ above $h$. For a certain leaf $y$ of $Y$, by Lemma \ref{lem:a1} (iii), there must exist a path $c$ from the root to $y$ which is vertex-disjoint with $N[Z]$. 
$c$ must pass $lca(Y)$ because $N[Y]$ does not contain any hybrid vertex in $\mathcal{H}$. 
So $lca(Y)$ is not in $N[Z]$ because $c$ is vertex-disjoint with $N[Z]$. $h'$ is neither in $N[Z]$ because $h' \in b(Z)$. So the path $lca(Y) \leadsto h'$ does not have common vertices with $N[Z]$. Hence, the subpath of $c$ from the root to $lca(Y)$  extended to $h'$ is vertex-disjoint with $N[Z]$. The later is a contradiction because $c_Z$ is a path $lca(Z) \hookrightarrow h$.

(ii) $lca(Y)$ is on $c_Z$. Similarly to the above case, there is a path $c$ from the root to $lca(Y)$ which is vertex disjoint with $N[Z]$. So $c$ intersects with $c_Z$ at a hybrid vertex $h'$ above $lca(Y)$.  The subpath of $c$ from the root to $h'$ is also vertex-disjoint with $N[Z]$.  The later is a contradiction because $c_Z$ is a path $lca(Z) \hookrightarrow h$.
\end{minipage}\hfill
\begin{minipage}[b]{5cm}
\includegraphics[scale=.5]{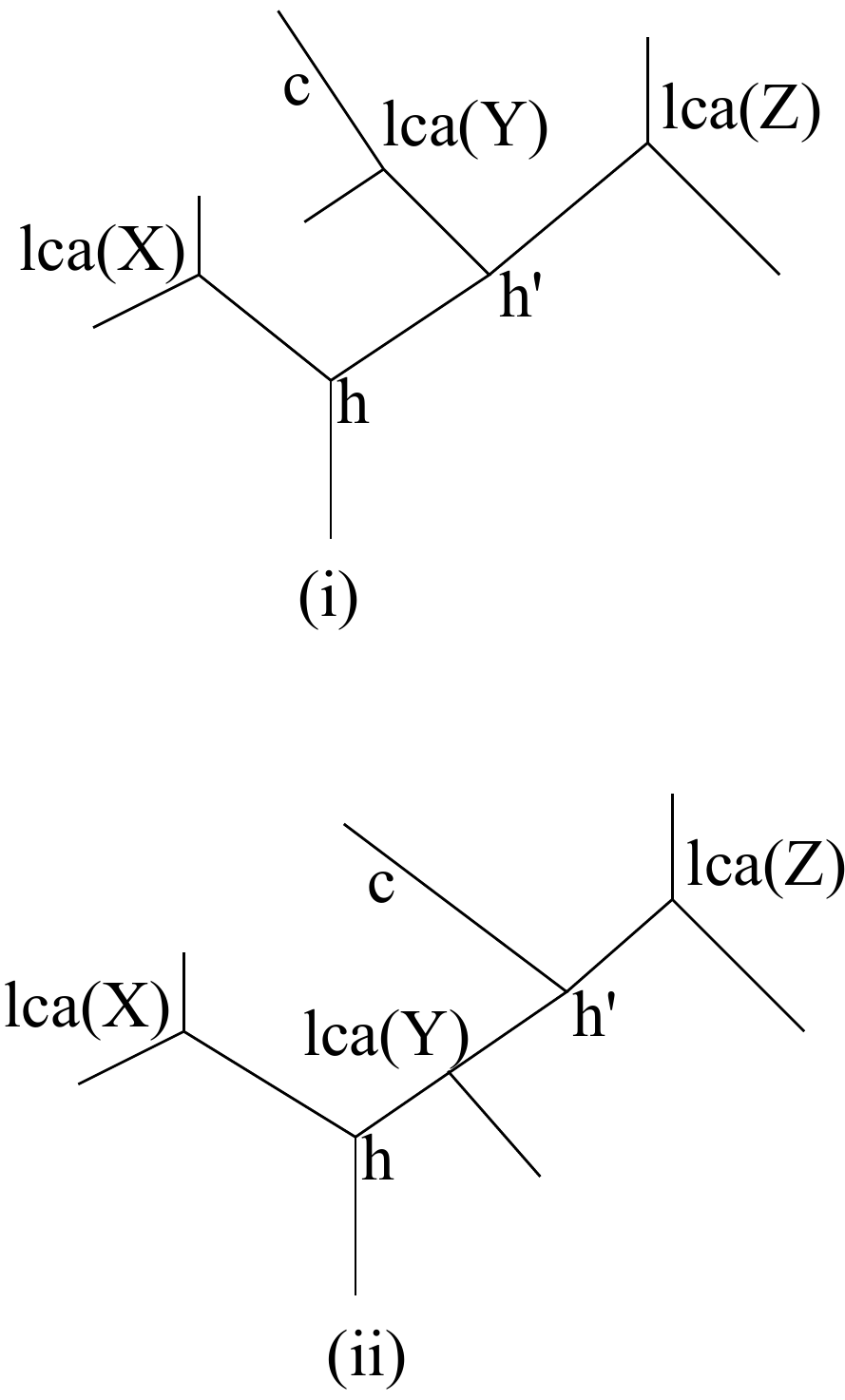}
\end{minipage}

(iii) Similarly for the case where $lca(Z)$ is on $c_Z$.
\end{pf}

\begin{lemma}
For any split SN-set $S$ of $N$ such that $a(S) = \emptyset$ and for any $x \not\in S$ which is below $lca(S)$, let $C_x$ be a path $lca(S) \leadsto x$, then $C_x$ contains one and only hybrid vertex in $b(S)$.
\label{lem:below}
\end{lemma}

\begin{pf}
By Lemma \ref{lem:a1} (i), $N[S]$ does not contain any hybrid vertex in $\mathcal{H}$. 
Since $x \not\in S$ then by Lemma \ref{lem:a1} (iii), there must exist a path $C'_x$ from the root to $x$ which is vertex-disjoint with $N[S]$. This path must intersect with $C_x$ at a hybrid vertex $h'$ above $x$. 
It means that there is at least one hybrid vertex on $C_x$ which has a path coming to it from outside of $N[S]$. Let $h_x$ be the highest hybrid vertex on $C_x$ having this property, then by definition $h_x \in b(S)$.
\end{pf}

\subsection{Restricting the searching class}
\label{sec:restrict}
We introduce here two lemmas \ref{lem:notsplit}, \ref{lem:notsplit1} which allow us to restrict the research to a class of level-$k$ phylogenetic networks having fewer split SN-sets without losing the ones having the minimum number of hybrid vertices. It is a generalisation to level $k$ of Theorem $3$ in \cite{IKKS08}.

For any split SN-set $S$ such that $a(S) = \emptyset$, let us define $\textbf{F}_\textbf{S}$ \index{$F_S$} to be the set of elements $(x,y,z)$
where $x,y$ are below $lca(S)$, $x,y \not\in S$, $z \in S$  such that $x|yz \in \mathcal{T}$.

\begin{lemma}
\label{lem:notsplit}
Given a level-$k$ network $N$ consistent with $\mathcal{T}$.  Let $S$ be an SN-set of $\mathcal{T}$ which is split in $N$ such that $a(S) = \emptyset$ and for any $(x,y,z) \in F_S$:

- either there is a path from the root to $x$ which is vertex-disjoint with a path from $lca(S)$ to $y$, 

- or there exist $2$ vertices $p,q$ in $N$ such that $p$ is above $lca(S)$ and there are $4$ internal vertex-disjoint paths $p \leadsto q$, $p \leadsto x$,  $q \leadsto y$, $q \leadsto z$.

Then, there is a level-$k$ network $N'$ consistent with $\mathcal{T}$, having the same number of hybrid vertices as $N$, in which $S$ is not split but is equal to a part of $P(N')$.
\end{lemma}

\begin{pf} Suppose that $S$ satisfies the condition stated in the lemma. Because $a(S) = \emptyset$, by Lemma \ref{lem:a1}, $S$ has only one lca. Let $G_S$ be the network obtained from $N[S]$ by contracting all arcs having one extremity of in-degree $1$ and out-degree $1$. So, $G_S$ has one lca, called $v_S$. We construct $N'$ from $N$ as follows (see the figure below in which only the part of the network that concerns $S$ is drawn):

\begin{minipage}[b]{6.5cm}
Delete from $N$ all subnetworks on $s_i$, contract all arcs having one extremity of in-degree $1$ and out-degree $1$,  add a new vertex $u_S$ in the middle of the arc coming to $lca(S)$ and then add a new arc from $u_S$ to $v_S$ where we attach $G_S$. So $(u_S,v_S)$ is a highest cut-arc of $N'$, i.e $S$ is a part of $P(N')$. 
\end{minipage}\hfill
\begin{minipage}[b]{9cm}
\def\svgwidth{9cm}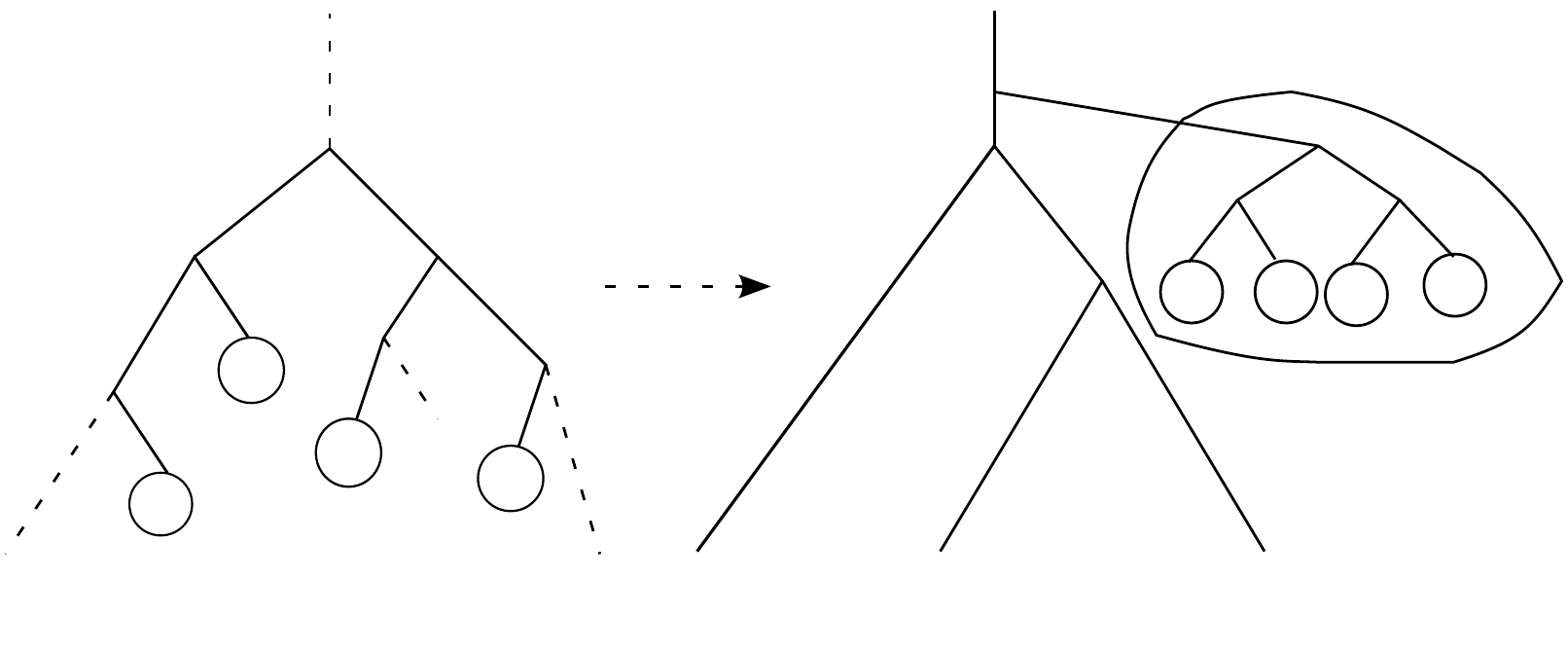
\end{minipage}

Because $a(S) = \emptyset$, there is not any hybrid vertex of $\mathcal{H}$ in $N[S]$. So, $G_S$ does not contain any hybrid vertex except those in the subnetworks on $s_i$. It implies that $N'$ has the same level and the same number of hybrid vertices as $N$. Now, it remains to show that $N'$ is consistent with all triplets of $\mathcal{T}$. For a triplet $x|yz \in \mathcal{T}$, we can distinguish the $6$ following cases:

(1) Since $S$ is an SN-set, the cases $x, y \in S$ and $z \notin S$ or $x, z \in S$ and $y \notin S$ are excluded.

\begin{figure}[ht]
\begin{center}
\includegraphics[scale=.7]{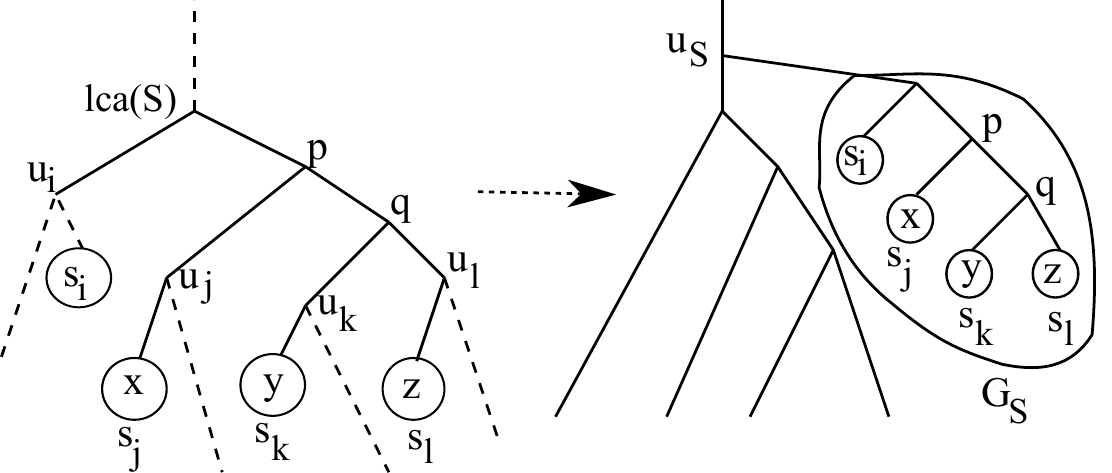}
  \caption{(2): $x,y,z \in S$}
\label{fig:xyz}
\end{center}
\end{figure}

(2) $x,y,z \in S$ (Figure \ref{fig:xyz}), so there exist $s_j,s_k,s_l$ such that $x \in s_j, y \in s_k, z \in s_l$ ($j,k,l$ are not necessarily distinct). By definition of consistency, $N$ has $2$ vertices $p,q$ and the internal vertex-disjoint paths $p \leadsto q, p \leadsto x, q \leadsto y, q \leadsto z$.
Because there is not any hybrid vertex in $N[S]$, any path from a vertex above $lca(S)$ to any leaf of $S$ must pass $lca(S)$. So, $p, q$ can not be above $lca(S)$ because otherwise, there are at least $2$ among the $4$ paths $p \leadsto q, p \leadsto x, q \leadsto y, q \leadsto z$ have $lca(S)$ as a common vertex. We deduce that $p, q$ are in $N[S]$. So, $x|yz$ is consistent with $N[S]$, or with $G_S$, and then is consistent with $N'$.

(3) $x,y,z \notin S$. We do not change the configuration of the network except the positions of the subnetworks on $s_i$. So all triplets of this case remain consistent with $N'$.

(4) $x \notin S, y,z \in S$. In $N'$, $y,z$ are below the highest cut-arc $(u_S,v_S)$ while $x$ is not. Hence, the triplet $x|yz$ is consistent with $N'$ in this case.

\begin{figure}[ht]
\begin{center}
\label{fig:iv}
  \subfigure[$lca(S)$ is below $p$ \label{fig:case1c}]{\includegraphics[scale=.7]{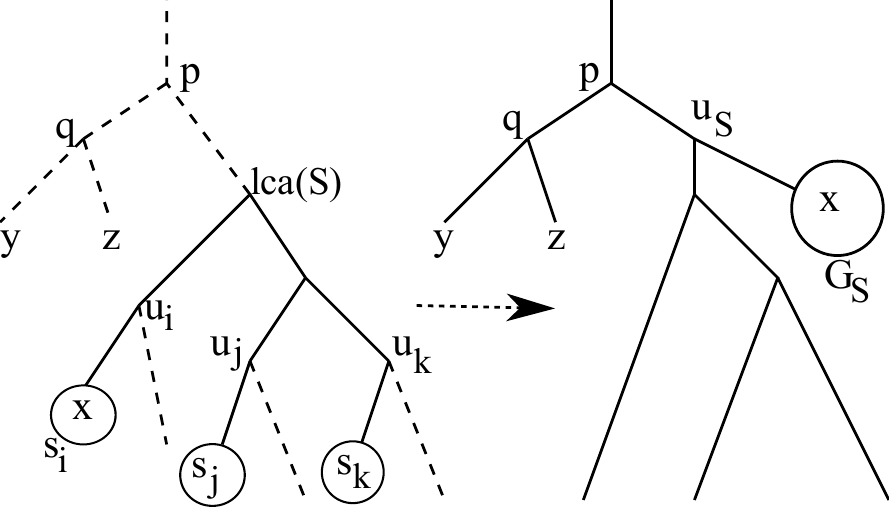}}\;\;\;\;\;\;
  \subfigure[$lca(S)$ is above $p$ \label{fig:case1a}]{\includegraphics[scale=.7]{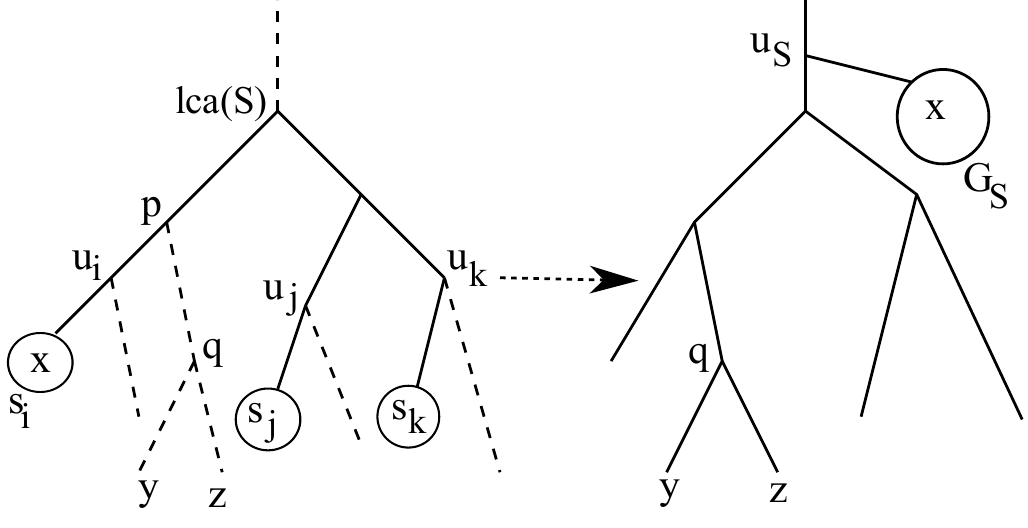}}
  \caption{(5): $x \in S, y,z \notin S$.}
  \label{fig:case4}
\end{center}
\end{figure}

(5) $x \in S, y,z \notin S$. Let $s_i$ be the child of $S$ such that $x \in s_i$. Let $p,q$ be two vertices of $N$ such that  there are internal vertex-disjoint paths $p \leadsto x$, $p \leadsto q$, $q \leadsto y$, $q \leadsto z$, we will prove that $lca(S)$ and $p$ are comparable. Suppose otherwise, then any path $p \leadsto u_i$ intersects with any path $lca(S) \leadsto u_i$ at a hybrid vertex $h'$. If $h' = u_i$ then $u_i$ is a hybrid vertex in $a(S)$, a contradiction because $a(S) = \emptyset$. So $h'$ is above $u_i$. However, $h'$ is below $lca(S)$, then $a(S)$ contains at least one vertex, a contradiction. Hence, we have two cases:

\hspace*{.5cm} a) If $lca(S)$ is below $p$ (see Figure \ref{fig:case1c}), then $q$ can not be below $lca(S)$ because otherwise, the paths $p \leadsto x$ and $p \leadsto q$ have $lca(S)$ as a common vertex, a contradiction. In $N$ the path $p \leadsto x$  must pass $lca(S)$. So the path $p \leadsto u_S$ in $N'$ is a subpath of $p \leadsto x$ in $N$. It implies that in $N'$, $p \leadsto u_S$ is also internal vertex-disjoint with $p \leadsto q$, $q \leadsto y$ and $q \leadsto z$. So, $x|yz$ is consistent with $N'$ because $x$ is below $u_S$.

\hspace*{.5cm} b) If $lca(S)$ is above or equal to $p$ (see Figure \ref{fig:case1a}), because $y,z$ are also below $lca(S)$ because they are below $p$. So, $y,z$ have a lca $q$ below $lca(S)$, i.e below $u_S$ in $N'$. Moreover, in $N'$, $x$ is hung below the highest cut-arc $(u_S,v_S)$, then any path $u_S \leadsto x$ is internal vertex-disjoint with $u_S \leadsto q, q \leadsto y, q \leadsto z$. In other words, $x|yz$ is consistent with $N'$.

\begin{figure}[ht]
\begin{center}
  \subfigure[$lca(S)$ is below $p$\label{fig:case2a}]{\includegraphics[scale=.7]{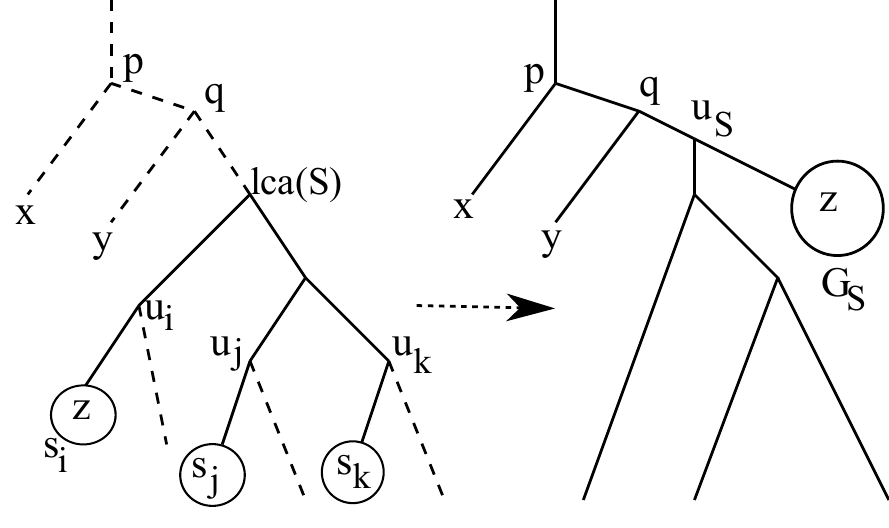}}\;\;\;\;\;\;\;
  \subfigure[$lca(S)$ is above $p$\label{fig:case2c}]{\includegraphics[scale=.7]{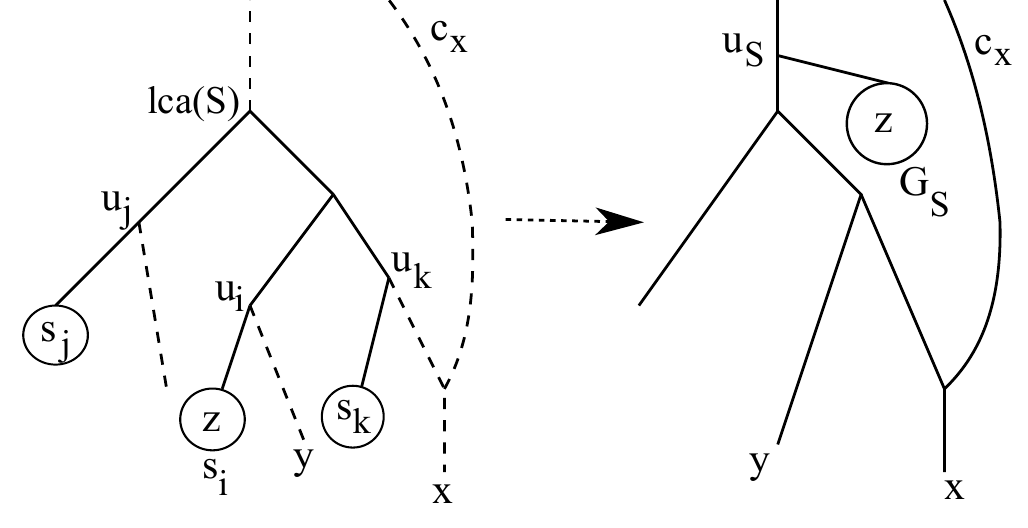}}
  \caption{(6): $x,y \notin S, z \in S$.}
  \label{fig:v}
\end{center}
\end{figure}

(6) $x,y \notin S, z \in S$. Let $s_i$ be the child of $S$ such that  $z \in s_i$. For any two vertices $p,q$ of $N$ such that there are $4$ internal vertex-disjoint paths $p \leadsto x, p \leadsto q, q \leadsto y, q \leadsto z$, by the same argument as that of the previous case (5), we deduce that $lca(S)$ and $p$ are comparable.

\hspace*{.5cm} a) If  $lca(S)$ is below $p$ (see Figure \ref{fig:case2a}), then in $N'$, $u_S$ is below $p$. Similarly to the case $(5a)$, the path $p \leadsto u_S$ in $N'$ is a subpath of $p \leadsto z$ in $N$, so it is internal vertex-disjoint with $p \leadsto x$, $q \leadsto y$ in $N'$. Because in $N'$, $z$ is below the cut-arc $(u_S,v_S)$, so every path $u_S \leadsto z$ is internal vertex-disjoint with the paths $p \leadsto x$, $p \leadsto q$, $q \leadsto y$, no matter where the position of $q$ is (above or below $lca(S)$). Hence, $x|yz$ is consistent with $N'$.

\hspace*{.5cm} b) If $lca(S)$ is above or equal to $p$ (see Figure \ref{fig:case2c}), then  $x,y$ are also below $lca(S)$ because  they are below $p$, then. It means that $(x,y,z) \in F_S$. According to the assumption:

- either there exist two other vertices $p',q'$ such that $p'$ is above $lca(S)$ and the internal vertex-disjoint path $p' \leadsto x$, $p' \leadsto q'$, $q' \leadsto y$, $q' \leadsto z$, then it returns to the case (a). 

- or there is a path $c_x$ from the root to $x$ which is vertex disjoint with a path $c_y$ from $lca(S)$ to $y$. In this case, $c_x$ in $N'$ is also vertex disjoint with the path from $u_S$ to $y$ by using $c_y$. So, $x|yz$ is also consistent with $N'$.

We conclude that $N'$ is consistent with all triplets of $\mathcal{T}$.
\end{pf}

For any split SN-set $S$ such that $a(S) = \emptyset$, let $F'_S$ be the subset of $F_S$ containing the elements $(x,y,z)$ such that:

(*) every path from the root to $x$ has common vertices with every path from $lca(S)$ to $y$, and 

(**) let $p,q$ be two vertices such that there are internal vertex disjoint paths $p \leadsto x$, $p \leadsto q$, $q \leadsto y$, $q \leadsto z$, then $p, q$ are in $N[S]$.

Therefore, $S$ does not satisfy the condition in Lemma \ref{lem:notsplit} iff $F'_S$ is not empty.

\begin{lemma}
Let $S$ be a split SN-set in $N$ such that $a(S) = \emptyset$ and $F'_S \neq \emptyset$. For any $(x,y,z) \in F'_S$, there exist two distinct hybrid vertices $h_x, h_y$ in $b(S)$ such that $h_x$ is above $x$, $h_y$ is above $y$, and $h_x$ is below $h_y$.
\label{lem:F'_S}
\end{lemma}

\begin{pf}
By (**), there is a vertex $p$ in $N[S]$ and two internal vertex disjoint paths $c_x: p \leadsto x$, and $c_y: p \leadsto y$. 
Let $C_x$ be the path from $lca(S)$ to $x$ containing $c_x$. Let $C_y$ be the path from $lca(S)$ to $y$ containing $c_y$. 
By Lemma \ref{lem:below}, there is a hybrid vertex $h_x$ of $b(S)$ on $C_x$, namely $h_x$. Because $h_x \not\in N[S]$, $p \in N[S]$, so $h_x$ is below $p$, i.e. $h_x$ is on $c_x$. Similarly, there is a hybrid vertex $h_y$ of $b(S)$ on $c_y$. $h_x \neq h_y$ because $c_x,c_y$ are vertex-disjoint.

We will prove that $h_x$ is below $h_y$. Let $c'_x$ be the subpath of $c_x$ from $h_x$ to $x$, and $c'_y$ be the subpath of $c_y$ from $h_y$ to $y$. Because $lca(S) \hookrightarrow h_x$, there is a path, called $C'$, from the root to $h_x$ which is vertex-disjoint with $N[S]$. 

\begin{minipage}[b]{7cm}
Let $C'_x$ be the path from the root to $x$ consisting of $C'$ and $c'_x$.
By (*), $C'_x$, $C_y$ must have common vertices. So, they must intersect at a hybrid vertex $h'$ because $C'_x$ does not pass $lca(S)$, while $x$ is below $lca(S)$. If $h'$ is below $h_y$, then $h_y$ is above $h_x$ because $h'$ is  above $h_x$. So  we are done. 
Suppose that $h'$ is above $h_y$. Because $h_y \in b(S)$, by definition every path from the root to $h'$ must have common vertices with $N[S]$. So, $C'$ has common vertices with $N[S]$, a contradiction.
\end{minipage}\hfill
\begin{minipage}[b]{8cm}
\includegraphics[scale=.65]{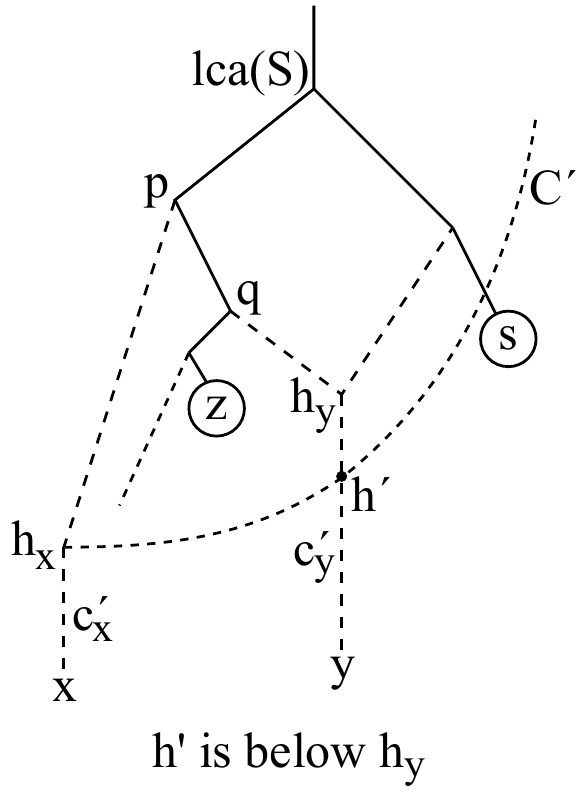}
\includegraphics[scale=.65]{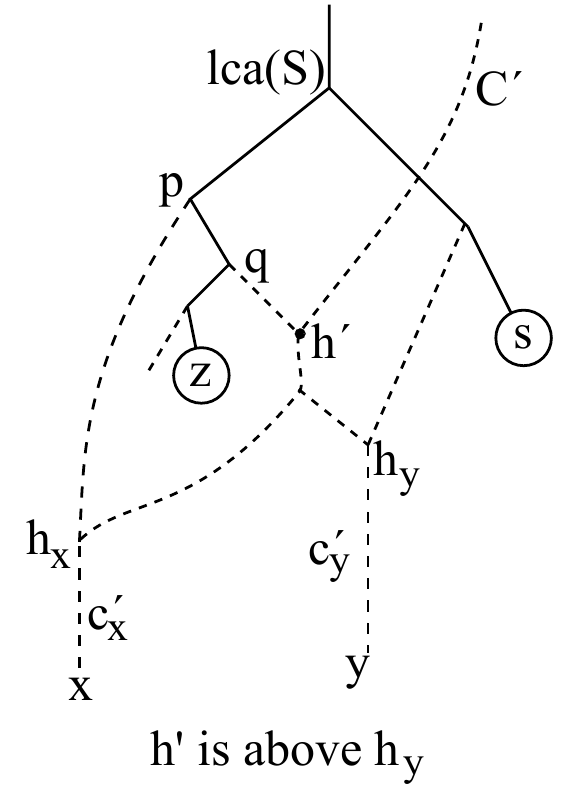}
\end{minipage}

Hence, $h_x$ is below $h_y$.
\end{pf}

\begin{lemma}
\label{lem:notsplit1}
Let $N$ be a level-$k$ network consistent with $\mathcal{T}$, let $S$ be an SN-set of $\mathcal{T}$ which is split in $N$ such that $a(S) = \emptyset$ and $|b(S)| \le 2$. Then there is a level-$k$ network $N'$ consistent with $\mathcal{T}$, having the same number of hybrid vertices as $N$, in which $S$ is not split but equal to a part of $P(N')$.
\end{lemma}

\begin{pf}
If $S$ satisfies the conditions in Lemma \ref{lem:notsplit}, then we are done. Suppose that $S$ does not satisfy the conditions in Lemma \ref{lem:notsplit}, so $F'_S$ is not empty. By using Lemma \ref{lem:F'_S} with a certain element $(x,y,z)$ of $F'_S$, we deduce that $b(S)$ contains at least $2$ hybrid vertices. So $|b(S)|=2$. Denote by $h_1,h_2$ the two hybrid vertices of $b(S)$. Also by Lemma \ref{lem:F'_S}, $h_1, h_2$ are \index{comparable} comparable. Suppose that $h_0$ is below $h_1$, then for any $(x,y,z) \in F'_S$, $h_x=h_1$, $h_y=h_2$, where $h_x, h_y$ are defined as in the proof of Lemma \ref{lem:F'_S}.

\begin{minipage}[b]{8cm}
We construct $G_S$ and modify $N$ by the same way that we did in the proof of Lemma \ref{lem:notsplit}. However, the position of $u_S$ below which we hang $G_S$ will be chosen differently. Let $p_0$ be a vertex of $N[S]$ such that there are two internal vertex disjoint paths $p_0 \leadsto h_1$ and $p_0 \leadsto h_2$. There exists always such a $p_0$, for example we can choose $p_0 = p$ which is defined in Lemma \ref{lem:F'_S} for a certain element $(x,y,z) \in F'_S$. $u_S$ is put in the middle of the arc going from $p$ on the path $p \leadsto h_2$. Denote the obtained network by $N'$. 
\end{minipage}\hfill
\begin{minipage}[b]{7cm}
\def\svgwidth{7cm}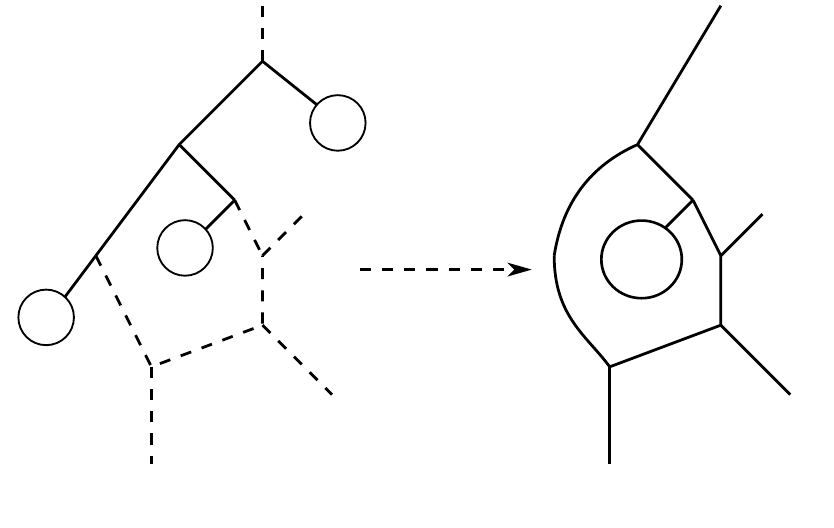
\end{minipage}

It is easy to see that $N'$ has the same level and the same number of hybrid vertices as $N$. We must check that all triplets $x|yz$ of $\mathcal{T}$ are consistent with $N'$. It can be verified that the proof of Lemma \ref{lem:notsplit} still holds here for all triplets except the cases $(5b)$ and $(6b)$.
 Let $p,q$ be two vertices of $N$ such that there exist $4$ internal vertex disjoint paths $p \leadsto q$, $p \leadsto x$, $q \leadsto y$, $q \leadsto z$.

$(5b)$ $x \in S, y,z \not\in S$ and $p$ is in $N[S]$, i.e. $y,z$ are below $lca(S)$. By Lemma \ref{lem:below}, any leaf below $lca(S)$ and not in $S$ must be below a hybrid vertex of $b(S)$. Moreover, $b(S)$ contains only $h_1, h_2$ and $h_1$ is below $h_2$. So both $y,z$ are below $h_2$. Hence, there exists a lca $q'$ of $y,z$ which is below $h_2$. Furthermore, $u_S$ is above $h_2$ then in $N'$ there are $4$ internal vertex-disjoint paths $u_S \leadsto x$, $u_S \leadsto q'$, $q' \leadsto y$ and $q' \leadsto z$, i.e $x|yz$ is consistent with $N'$ (Figure \ref{fig:case5b1}).

\begin{figure}[ht]
\begin{center}
  \subfigure[Case $(5b)$: $x\in S$, $y,z \not\in S$ \label{fig:case5b1}]{\def\svgwidth{6cm}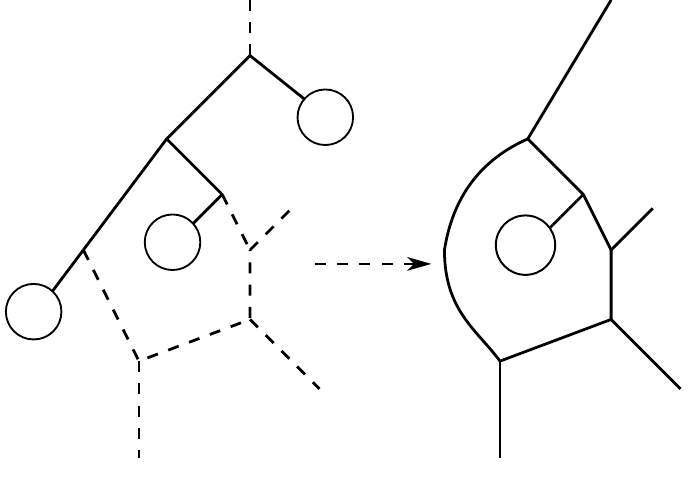}\;\; \;\;\;
  \subfigure[Case $(6b)$: $x,y \not\in S$, $z \in S$ \label{fig:case6b1}]{\def\svgwidth{6cm}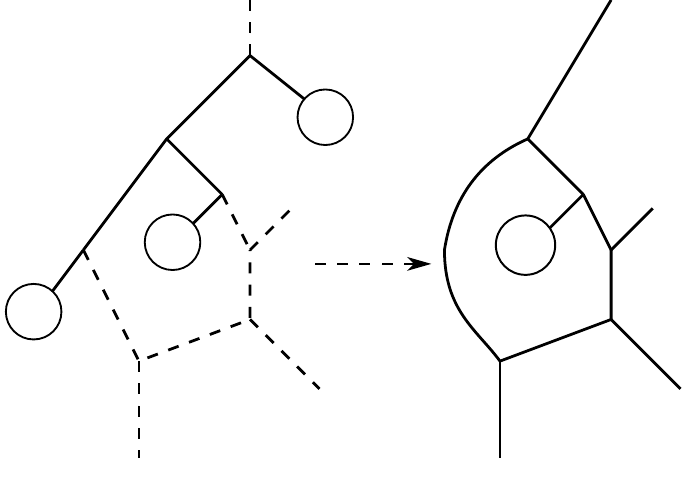}  
\caption{$x|yz$ is consistent with $N'$}
\label{fig:II_4b}
\end{center}
\end{figure}

$(6b)$ $x,y \not\in S$, $z \in S$ and $p$ is in $N[S]$, i.e. $x,y$ are below $lca(S)$. Then, each triplet $x|yz$ in this case corresponds to an element $(x,y,z)$ of $F_S$.

\hspace*{.5cm} - If $(x,y,z) \in F_S \setminus F'_S$, then it satisfies the properties in Lemma \ref{lem:notsplit}. We can use the same argument as that in the proof of Lemma \ref{lem:notsplit} to prove that $x|yz$ is consistent with $N'$.

\hspace*{.5cm} - Otherwise, $(x,y,z) \in F'_S$, so $h_1=h_x$, $h_2=h_y$. In other words, $x$ is below $h_1$ and $y$ is below $h_2$. By construction, $u_S$ is added on the path $p_0 \leadsto h_2$, so in $N'$ there are $4$ internal vertex-disjoint paths $p_0 \leadsto h_1 \leadsto x$, $p_0 \leadsto u_S$, $u_S \leadsto h_2 \leadsto y$ and $u_S \leadsto z$ (Figure \ref{fig:case6b1}). Hence, $x|yz$ is consistent with $N'$. 
\end{pf}

Using Lemmas \ref{lem:notsplit}, \ref{lem:notsplit1} without loss of the networks having the minimum number of hybrid vertices, we can restrict the research on the networks $N$ such that: each split SN-set $S$ of $N$ having $a(S) = \emptyset$ does not satisfy the $2$ conditions in these Lemmas. \textbf{In the following, we use only one of these two conditions, i.e. if $\textbf{a(S)} = \emptyset$, then $|\textbf{b(S)}| \ge \textbf{3}.$}

%% file: Figures/function_a.pdf_tex

\begingroup
  \makeatletter
  \providecommand\color[2][]{%
    \errmessage{(Inkscape) Color is used for the text in Inkscape, but the package 'color.sty' is not loaded}
    \renewcommand\color[2][]{}%
  }
  \providecommand\transparent[1]{%
    \errmessage{(Inkscape) Transparency is used (non-zero) for the text in Inkscape, but the package 'transparent.sty' is not loaded}
    \renewcommand\transparent[1]{}%
  }
  \providecommand\rotatebox[2]{#2}
  \ifx\svgwidth\undefined
    \setlength{\unitlength}{176pt}
  \else
    \setlength{\unitlength}{\svgwidth}
  \fi
  \global\let\svgwidth\undefined
  \makeatother
  \begin{picture}(1,1.22727273)%
    \put(0,0){\includegraphics[width=\unitlength]{Figures/function_a.pdf}}%
    \put(0.42355728,1.04568286){\color[rgb]{0,0,0}\makebox(0,0)[lb]{\smash{$lca(S)$}}}%
    \put(0.61164696,0.41970241){\color[rgb]{0,0,0}\makebox(0,0)[lb]{\smash{$h_2$}}}%
    \put(0.10700689,0.67401954){\color[rgb]{0,0,0}\makebox(0,0)[lb]{\smash{$h_0$}}}%
    \put(0.84346515,0.47879332){\color[rgb]{0,0,0}\makebox(0,0)[lb]{\smash{$u_1$}}}%
    \put(0.52073787,0.29697514){\color[rgb]{0,0,0}\makebox(0,0)[lb]{\smash{$u_2$}}}%
    \put(0.27528331,0.3515206){\color[rgb]{0,0,0}\makebox(0,0)[lb]{\smash{$u_3$}}}%
    \put(0.02528331,0.20606633){\color[rgb]{0,0,0}\makebox(0,0)[lb]{\smash{$u_4$}}}%
    \put(0.40044493,0.56648836){\color[rgb]{0,0,0}\makebox(0,0)[lb]{\smash{$u_3$}}}%
    \put(0.88437424,0.31970241){\color[rgb]{0,0,0}\makebox(0,0)[lb]{\smash{$s_1$}}}%
    \put(0.50255606,0.14242996){\color[rgb]{0,0,0}\makebox(0,0)[lb]{\smash{$s_2$}}}%
    \put(0.33437424,0.18788451){\color[rgb]{0,0,0}\makebox(0,0)[lb]{\smash{$s_3$}}}%
    \put(0.08437422,0.05152032){\color[rgb]{0,0,0}\makebox(0,0)[lb]{\smash{$s_4$}}}%
    \put(0.42528333,0.41515696){\color[rgb]{0,0,0}\makebox(0,0)[lb]{\smash{$s_3$}}}%
    \put(0.19264601,0.88729358){\color[rgb]{0,0,0}\makebox(0,0)[lb]{\smash{$h_1$}}}%
  \end{picture}%
\endgroup

%% file: Figures/Lemma_a.pdf_tex

\begingroup
  \makeatletter
  \providecommand\color[2][]{%
    \errmessage{(Inkscape) Color is used for the text in Inkscape, but the package 'color.sty' is not loaded}
    \renewcommand\color[2][]{}%
  }
  \providecommand\transparent[1]{%
    \errmessage{(Inkscape) Transparency is used (non-zero) for the text in Inkscape, but the package 'transparent.sty' is not loaded}
    \renewcommand\transparent[1]{}%
  }
  \providecommand\rotatebox[2]{#2}
  \ifx\svgwidth\undefined
    \setlength{\unitlength}{112pt}
  \else
    \setlength{\unitlength}{\svgwidth}
  \fi
  \global\let\svgwidth\undefined
  \makeatother
  \begin{picture}(1,1.07142857)%
    \put(0,0){\includegraphics[width=\unitlength]{Figures/Lemma_a.pdf}}%
    \put(0.20139557,0.82857664){\color[rgb]{0,0,0}\makebox(0,0)[lb]{\smash{$t_1$}}}%
    \put(0.72282415,0.82857664){\color[rgb]{0,0,0}\makebox(0,0)[lb]{\smash{$t_2$}}}%
    \put(0.05139556,0.55000521){\color[rgb]{0,0,0}\makebox(0,0)[lb]{\smash{$h_1$}}}%
    \put(0.85853844,0.54286236){\color[rgb]{0,0,0}\makebox(0,0)[lb]{\smash{$h_2$}}}%
    \put(0.02996699,0.33571906){\color[rgb]{0,0,0}\makebox(0,0)[lb]{\smash{$u_1$}}}%
    \put(0.85139558,0.35000478){\color[rgb]{0,0,0}\makebox(0,0)[lb]{\smash{$u_2$}}}%
    \put(0.23710984,0.09286236){\color[rgb]{0,0,0}\makebox(0,0)[lb]{\smash{$s_1$}}}%
    \put(0.63710987,0.10000521){\color[rgb]{0,0,0}\makebox(0,0)[lb]{\smash{$s_2$}}}%
  \end{picture}%
\endgroup

%% file: Figures/Lemma_a_.pdf_tex

\begingroup
  \makeatletter
  \providecommand\color[2][]{%
    \errmessage{(Inkscape) Color is used for the text in Inkscape, but the package 'color.sty' is not loaded}
    \renewcommand\color[2][]{}%
  }
  \providecommand\transparent[1]{%
    \errmessage{(Inkscape) Transparency is used (non-zero) for the text in Inkscape, but the package 'transparent.sty' is not loaded}
    \renewcommand\transparent[1]{}%
  }
  \providecommand\rotatebox[2]{#2}
  \ifx\svgwidth\undefined
    \setlength{\unitlength}{144pt}
  \else
    \setlength{\unitlength}{\svgwidth}
  \fi
  \global\let\svgwidth\undefined
  \makeatother
  \begin{picture}(1,0.83333333)%
    \put(0,0){\includegraphics[width=\unitlength]{Figures/Lemma_a_.pdf}}%
    \put(0.3189831,0.67595383){\color[rgb]{0,0,0}\makebox(0,0)[lb]{\smash{$p$}}}%
    \put(0.8467609,0.25373161){\color[rgb]{0,0,0}\makebox(0,0)[lb]{\smash{$u_j$}}}%
    \put(0.24676088,0.2426205){\color[rgb]{0,0,0}\makebox(0,0)[lb]{\smash{$u_i$}}}%
    \put(0.38009423,0.0592875){\color[rgb]{0,0,0}\makebox(0,0)[lb]{\smash{$s_i$}}}%
    \put(0.66898312,0.0592875){\color[rgb]{0,0,0}\makebox(0,0)[lb]{\smash{$s_j$}}}%
    \put(0.01898311,0.13706528){\color[rgb]{0,0,0}\makebox(0,0)[lb]{\smash{$x$}}}%
    \put(0.64120534,0.47595383){\color[rgb]{0,0,0}\makebox(0,0)[lb]{\smash{$lca(S)=q$}}}%
  \end{picture}%
\endgroup

%% file: Figures/function_b.pdf_tex

\begingroup
  \makeatletter
  \providecommand\color[2][]{%
    \errmessage{(Inkscape) Color is used for the text in Inkscape, but the package 'color.sty' is not loaded}
    \renewcommand\color[2][]{}%
  }
  \providecommand\transparent[1]{%
    \errmessage{(Inkscape) Transparency is used (non-zero) for the text in Inkscape, but the package 'transparent.sty' is not loaded}
    \renewcommand\transparent[1]{}%
  }
  \providecommand\rotatebox[2]{#2}
  \ifx\svgwidth\undefined
    \setlength{\unitlength}{216pt}
  \else
    \setlength{\unitlength}{\svgwidth}
  \fi
  \global\let\svgwidth\undefined
  \makeatother
  \begin{picture}(1,0.85185185)%
    \put(0,0){\includegraphics[width=\unitlength]{Figures/function_b.pdf}}%
    \put(0.23703704,0.72962962){\color[rgb]{0,0,0}\makebox(0,0)[lb]{\smash{$lca(S)$}}}%
    \put(0.67037037,0.25555555){\color[rgb]{0,0,0}\makebox(0,0)[lb]{\smash{$h_0$}}}%
    \put(0.79259237,0.12222221){\color[rgb]{0,0,0}\makebox(0,0)[lb]{\smash{$h_1$}}}%
    \put(0.79044721,0.47740545){\color[rgb]{0,0,0}\makebox(0,0)[lb]{\smash{$u_1$}}}%
    \put(0.55925926,0.67777777){\color[rgb]{0,0,0}\makebox(0,0)[lb]{\smash{$u_2$}}}%
    \put(0.33333333,0.36666666){\color[rgb]{0,0,0}\makebox(0,0)[lb]{\smash{$u_4$}}}%
    \put(0.51637325,0.39592397){\color[rgb]{0,0,0}\makebox(0,0)[lb]{\smash{$u_3$}}}%
    \put(0.0867435,0.4255536){\color[rgb]{0,0,0}\makebox(0,0)[lb]{\smash{$h_2$}}}%
    \put(0.8259257,0.35185184){\color[rgb]{0,0,0}\makebox(0,0)[lb]{\smash{$s_1$}}}%
    \put(0.46296296,0.48888888){\color[rgb]{0,0,0}\makebox(0,0)[lb]{\smash{$s_2$}}}%
    \put(0.48148148,0.23703703){\color[rgb]{0,0,0}\makebox(0,0)[lb]{\smash{$s_3$}}}%
    \put(0.32962963,0.19999999){\color[rgb]{0,0,0}\makebox(0,0)[lb]{\smash{$s_4$}}}%
    \put(0.00785674,0.28582062){\color[rgb]{0,0,0}\makebox(0,0)[lb]{\smash{$h_3$}}}%
  \end{picture}%
\endgroup

%% file: Figures/modify.pdf_tex

\begingroup
  \makeatletter
  \providecommand\color[2][]{%
    \errmessage{(Inkscape) Color is used for the text in Inkscape, but the package 'color.sty' is not loaded}
    \renewcommand\color[2][]{}%
  }
  \providecommand\transparent[1]{%
    \errmessage{(Inkscape) Transparency is used (non-zero) for the text in Inkscape, but the package 'transparent.sty' is not loaded}
    \renewcommand\transparent[1]{}%
  }
  \providecommand\rotatebox[2]{#2}
  \ifx\svgwidth\undefined
    \setlength{\unitlength}{464pt}
  \else
    \setlength{\unitlength}{\svgwidth}
  \fi
  \global\let\svgwidth\undefined
  \makeatother
  \begin{picture}(1,0.4137931)%
    \put(0,0){\includegraphics[width=\unitlength]{Figures/modify.pdf}}%
    \put(0.21896552,0.31724137){\color[rgb]{0,0,0}\makebox(0,0)[lb]{\smash{$lca(S)$}}}%
    \put(0.02758621,0.16034482){\color[rgb]{0,0,0}\makebox(0,0)[lb]{\smash{$u_2$}}}%
    \put(0.08448276,0.25){\color[rgb]{0,0,0}\makebox(0,0)[lb]{\smash{$u_1$}}}%
    \put(0.25,0.18793103){\color[rgb]{0,0,0}\makebox(0,0)[lb]{\smash{$u_3$}}}%
    \put(0.35689655,0.1775862){\color[rgb]{0,0,0}\makebox(0,0)[lb]{\smash{$u_4$}}}%
    \put(0.14482759,0.17068965){\color[rgb]{0,0,0}\makebox(0,0)[lb]{\smash{$s_1$}}}%
    \put(0.0862069,0.08620689){\color[rgb]{0,0,0}\makebox(0,0)[lb]{\smash{$s_2$}}}%
    \put(0.20517241,0.11724137){\color[rgb]{0,0,0}\makebox(0,0)[lb]{\smash{$s_3$}}}%
    \put(0.30862069,0.10344827){\color[rgb]{0,0,0}\makebox(0,0)[lb]{\smash{$s_4$}}}%
    \put(0.74482759,0.22068965){\color[rgb]{0,0,0}\makebox(0,0)[lb]{\smash{$s_1$}}}%
    \put(0.80172414,0.22241379){\color[rgb]{0,0,0}\makebox(0,0)[lb]{\smash{$s_2$}}}%
    \put(0.84655172,0.21896551){\color[rgb]{0,0,0}\makebox(0,0)[lb]{\smash{$s_3$}}}%
    \put(0.91206897,0.22586206){\color[rgb]{0,0,0}\makebox(0,0)[lb]{\smash{$s_4$}}}%
    \put(0.83103448,0.15172413){\color[rgb]{0,0,0}\makebox(0,0)[lb]{\smash{$G_S$}}}%
    \put(0.57241379,0.35344827){\color[rgb]{0,0,0}\makebox(0,0)[lb]{\smash{$u_S$}}}%
    \put(0.83103448,0.32413793){\color[rgb]{0,0,0}\makebox(0,0)[lb]{\smash{$v_S$}}}%
    \put(0.33275862,0.01034482){\color[rgb]{0,0,0}\makebox(0,0)[lb]{\smash{Modify $N \rightarrow N'$}}}%
  \end{picture}%
\endgroup

%% file: Figures/modify1.pdf_tex

\begingroup
  \makeatletter
  \providecommand\color[2][]{%
    \errmessage{(Inkscape) Color is used for the text in Inkscape, but the package 'color.sty' is not loaded}
    \renewcommand\color[2][]{}%
  }
  \providecommand\transparent[1]{%
    \errmessage{(Inkscape) Transparency is used (non-zero) for the text in Inkscape, but the package 'transparent.sty' is not loaded}
    \renewcommand\transparent[1]{}%
  }
  \providecommand\rotatebox[2]{#2}
  \ifx\svgwidth\undefined
    \setlength{\unitlength}{240pt}
  \else
    \setlength{\unitlength}{\svgwidth}
  \fi
  \global\let\svgwidth\undefined
  \makeatother
  \begin{picture}(1,0.63333333)%
    \put(0,0){\includegraphics[width=\unitlength]{Figures/modify1.pdf}}%
    \put(0.18334489,0.15959085){\color[rgb]{0,0,0}\makebox(0,0)[lb]{\smash{$h_1$}}}%
    \put(0.31746956,0.2985205){\color[rgb]{0,0,0}\makebox(0,0)[lb]{\smash{$h_2$}}}%
    \put(0.31823331,0.56233682){\color[rgb]{0,0,0}\makebox(0,0)[lb]{\smash{$lca(S)$}}}%
    \put(0.15834874,0.47062987){\color[rgb]{0,0,0}\makebox(0,0)[lb]{\smash{$p_0$}}}%
    \put(0.0271617,0.24209431){\color[rgb]{0,0,0}\makebox(0,0)[lb]{\smash{$s_3$}}}%
    \put(0.37821465,0.47437377){\color[rgb]{0,0,0}\makebox(0,0)[lb]{\smash{$s_1$}}}%
    \put(0.19304494,0.32430745){\color[rgb]{0,0,0}\makebox(0,0)[lb]{\smash{$s_2$}}}%
    \put(0.73494258,0.16125162){\color[rgb]{0,0,0}\makebox(0,0)[lb]{\smash{$h_1$}}}%
    \put(0.86753194,0.2985205){\color[rgb]{0,0,0}\makebox(0,0)[lb]{\smash{$h_2$}}}%
    \put(0.70675095,0.46903218){\color[rgb]{0,0,0}\makebox(0,0)[lb]{\smash{$p_0$}}}%
    \put(0.72540731,0.30801798){\color[rgb]{0,0,0}\makebox(0,0)[lb]{\smash{$G_S$}}}%
    \put(0.82872833,0.40425435){\color[rgb]{0,0,0}\makebox(0,0)[lb]{\smash{$u_S$}}}%
    \put(0.28342758,0.39946105){\color[rgb]{0,0,0}\makebox(0,0)[lb]{\smash{$u_2$}}}%
    \put(0.05335238,0.33235574){\color[rgb]{0,0,0}\makebox(0,0)[lb]{\smash{$u_3$}}}%
    \put(0.38681641,0.00996093){\color[rgb]{0,0,0}\makebox(0,0)[lb]{\smash{Modify $N \rightarrow N'$}}}%
  \end{picture}%
\endgroup

%% file: Figures/case5b1.pdf_tex

\begingroup
  \makeatletter
  \providecommand\color[2][]{%
    \errmessage{(Inkscape) Color is used for the text in Inkscape, but the package 'color.sty' is not loaded}
    \renewcommand\color[2][]{}%
  }
  \providecommand\transparent[1]{%
    \errmessage{(Inkscape) Transparency is used (non-zero) for the text in Inkscape, but the package 'transparent.sty' is not loaded}
    \renewcommand\transparent[1]{}%
  }
  \providecommand\rotatebox[2]{#2}
  \ifx\svgwidth\undefined
    \setlength{\unitlength}{200pt}
  \else
    \setlength{\unitlength}{\svgwidth}
  \fi
  \global\let\svgwidth\undefined
  \makeatother
  \begin{picture}(1,0.7)%
    \put(0,0){\includegraphics[width=\unitlength]{Figures/case5b1.pdf}}%
    \put(0.12431075,0.15365795){\color[rgb]{0,0,0}\makebox(0,0)[lb]{\smash{$h_1$}}}%
    \put(0.36694003,0.30841991){\color[rgb]{0,0,0}\makebox(0,0)[lb]{\smash{$h_2$}}}%
    \put(0.36387216,0.62300731){\color[rgb]{0,0,0}\makebox(0,0)[lb]{\smash{$lca(S)$}}}%
    \put(0.17201068,0.51295897){\color[rgb]{0,0,0}\makebox(0,0)[lb]{\smash{$p_0$}}}%
    \put(0.17142857,0.00266942){\color[rgb]{0,0,0}\makebox(0,0)[lb]{\smash{$y$}}}%
    \put(0.43984964,0.10236864){\color[rgb]{0,0,0}\makebox(0,0)[lb]{\smash{$z$}}}%
    \put(0.01458623,0.2387163){\color[rgb]{0,0,0}\makebox(0,0)[lb]{\smash{$s_3$}}}%
    \put(0.43983411,0.51745165){\color[rgb]{0,0,0}\makebox(0,0)[lb]{\smash{$x$}}}%
    \put(0.21364612,0.33737206){\color[rgb]{0,0,0}\makebox(0,0)[lb]{\smash{$s_2$}}}%
    \put(0.64622797,0.15166601){\color[rgb]{0,0,0}\makebox(0,0)[lb]{\smash{$h_1$}}}%
    \put(0.88502271,0.30841991){\color[rgb]{0,0,0}\makebox(0,0)[lb]{\smash{$h_2$}}}%
    \put(0.69009332,0.51104174){\color[rgb]{0,0,0}\makebox(0,0)[lb]{\smash{$p_0$}}}%
    \put(0.69718042,0.00266942){\color[rgb]{0,0,0}\makebox(0,0)[lb]{\smash{$y$}}}%
    \put(0.95409772,0.10428612){\color[rgb]{0,0,0}\makebox(0,0)[lb]{\smash{$z$}}}%
    \put(0.69056689,0.25805907){\color[rgb]{0,0,0}\makebox(0,0)[lb]{\smash{$G_S$}}}%
    \put(0.834474,0.43330834){\color[rgb]{0,0,0}\makebox(0,0)[lb]{\smash{$u_S$}}}%
    \put(0.32210529,0.42755639){\color[rgb]{0,0,0}\makebox(0,0)[lb]{\smash{$u_2$}}}%
    \put(0.04601504,0.34703002){\color[rgb]{0,0,0}\makebox(0,0)[lb]{\smash{$u_3$}}}%
    \put(0.72914062,0.33070311){\color[rgb]{0,0,0}\makebox(0,0)[lb]{\smash{$x$}}}%
    \put(0.84667969,0.1892578){\color[rgb]{0,0,0}\makebox(0,0)[lb]{\smash{$q'$}}}%
    \put(0.33615234,0.19509398){\color[rgb]{0,0,0}\makebox(0,0)[lb]{\smash{$q'$}}}%
    \put(0.44663672,0.45611718){\color[rgb]{0,0,0}\makebox(0,0)[lb]{\smash{$s_1$}}}%
  \end{picture}%
\endgroup

%% file: Figures/case6b1.pdf_tex

\begingroup
  \makeatletter
  \providecommand\color[2][]{%
    \errmessage{(Inkscape) Color is used for the text in Inkscape, but the package 'color.sty' is not loaded}
    \renewcommand\color[2][]{}%
  }
  \providecommand\transparent[1]{%
    \errmessage{(Inkscape) Transparency is used (non-zero) for the text in Inkscape, but the package 'transparent.sty' is not loaded}
    \renewcommand\transparent[1]{}%
  }
  \providecommand\rotatebox[2]{#2}
  \ifx\svgwidth\undefined
    \setlength{\unitlength}{200pt}
  \else
    \setlength{\unitlength}{\svgwidth}
  \fi
  \global\let\svgwidth\undefined
  \makeatother
  \begin{picture}(1,0.7)%
    \put(0,0){\includegraphics[width=\unitlength]{Figures/case6b1.pdf}}%
    \put(0.12431075,0.15365795){\color[rgb]{0,0,0}\makebox(0,0)[lb]{\smash{$h_1$}}}%
    \put(0.36694003,0.30841991){\color[rgb]{0,0,0}\makebox(0,0)[lb]{\smash{$h_2$}}}%
    \put(0.36387216,0.62300731){\color[rgb]{0,0,0}\makebox(0,0)[lb]{\smash{$lca(S)$}}}%
    \put(0.17201068,0.51295897){\color[rgb]{0,0,0}\makebox(0,0)[lb]{\smash{$p_0$}}}%
    \put(0.17142857,0.00266942){\color[rgb]{0,0,0}\makebox(0,0)[lb]{\smash{$x$}}}%
    \put(0.43984964,0.10236864){\color[rgb]{0,0,0}\makebox(0,0)[lb]{\smash{$y$}}}%
    \put(0.01458623,0.2387163){\color[rgb]{0,0,0}\makebox(0,0)[lb]{\smash{$s_3$}}}%
    \put(0.21962268,0.33737206){\color[rgb]{0,0,0}\makebox(0,0)[lb]{\smash{$z$}}}%
    \put(0.64622797,0.15166601){\color[rgb]{0,0,0}\makebox(0,0)[lb]{\smash{$h_1$}}}%
    \put(0.88502271,0.30841991){\color[rgb]{0,0,0}\makebox(0,0)[lb]{\smash{$h_2$}}}%
    \put(0.69009332,0.51104174){\color[rgb]{0,0,0}\makebox(0,0)[lb]{\smash{$p_0$}}}%
    \put(0.69718042,0.00266942){\color[rgb]{0,0,0}\makebox(0,0)[lb]{\smash{$x$}}}%
    \put(0.95409772,0.10428612){\color[rgb]{0,0,0}\makebox(0,0)[lb]{\smash{$y$}}}%
    \put(0.69056689,0.25805907){\color[rgb]{0,0,0}\makebox(0,0)[lb]{\smash{$G_S$}}}%
    \put(0.834474,0.43330834){\color[rgb]{0,0,0}\makebox(0,0)[lb]{\smash{$u_S$}}}%
    \put(0.32210529,0.42755639){\color[rgb]{0,0,0}\makebox(0,0)[lb]{\smash{$u_2$}}}%
    \put(0.04601504,0.34703002){\color[rgb]{0,0,0}\makebox(0,0)[lb]{\smash{$u_3$}}}%
    \put(0.733125,0.3326953){\color[rgb]{0,0,0}\makebox(0,0)[lb]{\smash{$z$}}}%
    \put(0.84667969,0.1892578){\color[rgb]{0,0,0}\makebox(0,0)[lb]{\smash{$q'$}}}%
    \put(0.33615234,0.19509398){\color[rgb]{0,0,0}\makebox(0,0)[lb]{\smash{$q'$}}}%
    \put(0.43866797,0.51389061){\color[rgb]{0,0,0}\makebox(0,0)[lb]{\smash{$s_1$}}}%
    \put(0.18765233,0.2768203){\color[rgb]{0,0,0}\makebox(0,0)[lb]{\smash{$s_2$}}}%
  \end{picture}%
\endgroup

%% file: Bound.tex
\section{A bound on the restricted networks class}
\label{sec:strict_bound}
As concluded in Section \ref{sec:restrict}, without loss the level-$k$ networks having the minimum number of hybrid vertices, we can suppose that the constructing level-$k$ networks $N$ having the following property: for any split SN-set $S$ of $N$, if $a(S) = \emptyset$, then $|b(S)| \ge 3$.

Let $\mathcal{S}$ be the set of SN-sets of $\mathcal{T}$ that are split in $N$. We will bound $|\mathcal{S}|$ by a stricter linear function of $k$.
To this aim, the functions $a,b$ defined in Section \ref{sec:split} and another function $t$ defined in the next will be explored.
We will introduce some lemmas showing some properties of each function which allow us to establish the relation between the number of hybrid vertices in $\mathcal{H}$ and the cardinality of $\mathcal{S}$.

\subsection{Partition $\mathcal{H}$ and $\mathcal{S}$ by the functions $a,b$}

\begin{figure}[ht]
\begin{center}
\def\svgwidth{\textwidth}
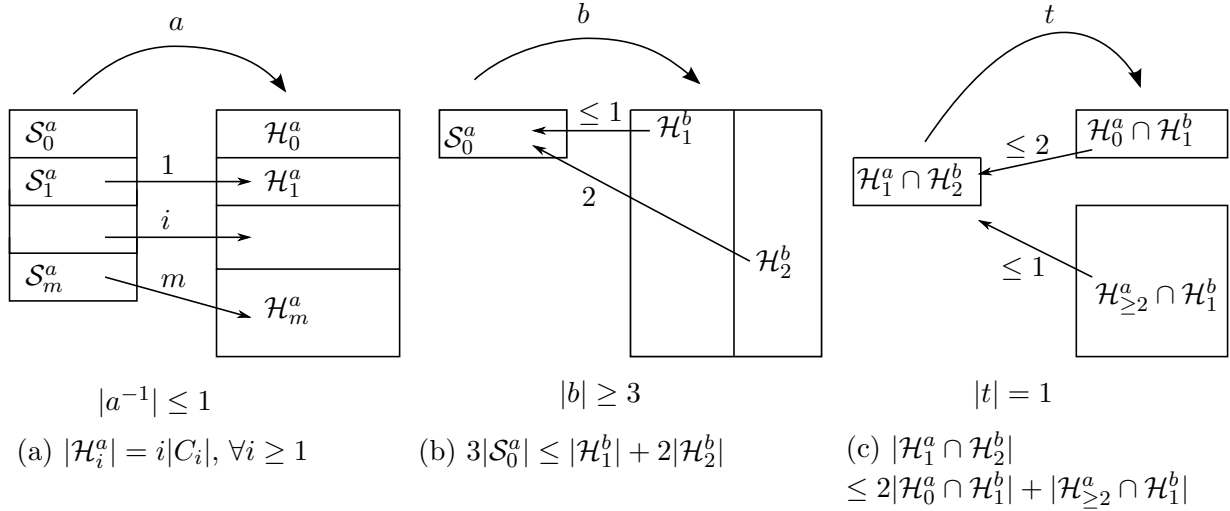
\caption{The $3$ functions $a,b,t$ and their properties. The set of hybrid vertices: $\mathcal{H} = \cup~\mathcal{H}^a_i = \mathcal{H}^b_1 \cup~\mathcal{H}^b_2$. The set of split SN-sets: $\mathcal{S}=\cup~\mathcal{S}^a_i$.}
\label{fig:set}
\end{center}
\end{figure}

Using the function $a$, we partition $\mathcal{H}$ and $\mathcal{S}$ into several subsets: (Figure \ref{fig:set}(a)).

- $\mathcal{H}^a_0=\{h\in \mathcal{H}| a^{-1}(h) = \emptyset\}$, \index{$\mathcal{H}^a_0$} and $\mathcal{S}^a_0 = \{S \in \mathcal{S}| a(S) = \emptyset\}$. \index{$\mathcal{S}^a_0$} 

- For any $i \ge 1$, $\mathcal{S}^a_i = \{S \in \mathcal{S}|~|a(S)|=i\}$, so all $\mathcal{S}^a_i$ are pairwise disjoint. \index{$\mathcal{S}^a_i$, $i \ge 1$}

- $\mathcal{H}^a_i$ \index{$\mathcal{H}^a_i$, $i \ge 1$} is the image of $\mathcal{S}^a_i$ by the function $a$. By Lemma \ref{lem:function}, all $\mathcal{H}^a_i$ are pairwise disjoint. 

\begin{lemma}
\label{lem:1}
$|\mathcal{S}| \le  k + |\mathcal{S}^a_0| - |\mathcal{H}^a_0| - \frac{1}{2}|\mathcal{H}^a_{\ge 2}|$
\end{lemma}

\begin{pf}
By definition, all $\mathcal{H}^a_i$ are pairwise disjoint, and $|\mathcal{H}^a_i|=i|\mathcal{S}^a_i|$ for any $i \ge 1$.

Then: $|\mathcal{S}| = |\mathcal{S}^a_0| + |\mathcal{S}^a_1| + \sum_{i \ge 2} |\mathcal{S}^a_i| =|\mathcal{S}^a_0| + |\mathcal{H}^a_1| + \sum_{i \ge 2} \frac{1}{i}|\mathcal{H}^a_i|$.

Furthermore, $|\mathcal{H}| = |\mathcal{H}^a_0|+ |\mathcal{H}^a_1| + \sum_{i \ge 2}|\mathcal{H}^a_i| \le k$

So, $|\mathcal{S}| \le k + |\mathcal{S}^a_0| - |\mathcal{H}^a_0| - \sum_{i \ge 2}\frac{i-1}{i}|\mathcal{H}^a_i| \le k + |\mathcal{S}^a_0| - |\mathcal{H}^a_0| - \frac{1}{2}|\mathcal{H}^a_{\ge 2}|$
\end{pf}

Then, in order to bound $|\mathcal{S}|$, it remains to determine the relations between $|\mathcal{S}^a_0|$,  $|\mathcal{H}^a_0|$ and $|\mathcal{H}^a_{\ge 2}|$. 

Due to Lemma \ref{lem:function_b}, we can use the function $b$ to partition $\mathcal{H}$ into $2$ subsets (Figure \ref{fig:set}(b)):

$\mathcal{H}^b_1 = \{h \in \mathcal{H}|$ there is at most one split SN-set $X$ of $\mathcal{S}^a_0$ such that $h \in b(X)\}$.

$\mathcal{H}^b_2 = \{h \in \mathcal{H}|$ there  are two split SN-sets $X,Y$ of $\mathcal{S}^a_0$ such that $h \in b(X) \cap b(Y)\}$.

So, $\mathcal{H}^b_1$ and $\mathcal{H}^b_2$ are disjoint. 

\begin{lemma}
\label{lem:2}
(i) $3|\mathcal{S}^a_0| \le |\mathcal{H}^b_1| +2|\mathcal{H}^b_2|$.

(ii) $|\mathcal{S}| \le \frac{4}{3}k + \frac{1}{3}|\mathcal{H}^b_2| - |\mathcal{H}^a_0| - \frac{1}{2}|\mathcal{H}^a_{\ge 2}|$
\end{lemma}

\begin{pf}
(i) According to the assumption on restricted searching class, $\forall S \in \mathcal{S}^a_0$, $|b(S)| \ge 3$. With the definition of $\mathcal{H}^b_1, \mathcal{H}^b_2$ as above, we are done (see Figure \ref{fig:set}(b)).

(ii) By Lemma \ref{lem:1} and Claim (i), we have:

$|\mathcal{S}| \le k + |\mathcal{S}^a_0| - |\mathcal{H}^a_0| - \frac{1}{2}|\mathcal{H}^a_{\ge 2}|  \le k + \frac{1}{3}|\mathcal{H}^b_1|+\frac{2}{3}|\mathcal{H}^b_2| - |\mathcal{H}^a_0| - \frac{1}{2}|\mathcal{H}^a_{\ge 2}|$. 

We have $|\mathcal{H}^b_1|+|\mathcal{H}^b_2| = |\mathcal{H}| \le  k$, then $|\mathcal{H}^b_1| \le k - |\mathcal{H}^b_2|$.

So, $|\mathcal{S}| \le \frac{4}{3}k + \frac{1}{3}|\mathcal{H}^b_2| - |\mathcal{H}^a_0| - \frac{1}{2}|\mathcal{H}^a_{\ge 2}|$.
\end{pf}

To reach our main result we need to find the relation between  $|\mathcal{H}^b_2|$ and $|\mathcal{H}^a_0|$, $|\mathcal{H}^a_{\ge 2}|$. To this aim, a function $t$ is introduced.

\subsection{Function $t$}
\label{sec:function_t}
For any $h \in \mathcal{H}^a_{\ge 1}$, denote by $\textbf{S}_\textbf{h}$ the only split SN-set such that $h \in a(S_h)$.
For any $h$ in $\mathcal{H}^a_1$, denote by $\textbf{P}_{\textbf{h}}$ \index{$P_h$} a path from $lca(S_h)$ to $h$. We can always choose for each $h$ in  $\mathcal{H}^a_1$ a path $P_h$ such that: \textit{$\forall h_1, h_2 \in \mathcal{H}^a_1$, if $P_{h_1}$, $P_{h_2}$ have to common vertices $u,v$ such that $u$ is above $v$, then the two subpaths of $P_{h_1}$ and $P_{h_2}$ from $u$ to $v$ are the same.} It is easy to see that there exists always such a path for each hybrid vertex of $\mathcal{H}^a_1$. Indeed, if the two subpaths of $P_{h_1}$ and $P_{h_2}$ from $u$ to $v$ are not the same, then we need only to change the subpath of $P_{h_2}$ from $u$ to $v$  to be the same as the subpath of $P_{h_1}$ from $u$ to $v$. The new path $P_{h_2}$ is always a path from $lca(S_{h_2})$ to $h_2$.

$\forall h \in \mathcal{H}^a_1$, we define some sets of hybrid vertices associated with $h$ as follows:

$I_0(h)$ \index{$I_0(h)$} is the set of hybrid vertices in $\mathcal{H}^a_0$ different from $h$ on $P_h$.

\begin{minipage}[b]{10.5cm}
$I_1(h)$ \index{$I_1(h)$} is the set of hybrid vertices $h'$ in $\mathcal{H}^a_1$ different from $h$ on $P_h$ such that $P_h$ and $P_{h'}$ have common vertices above $h$. 

$I_2(h)$ \index{$I_1(h)$} is the set of hybrid vertices $h'$ in $\mathcal{H}^a_{\ge 2}$ different from $h$ on $P_h$ such that $P_h$ and $ N[S_{h'}]$ have common vertices above $h$.

Finally, $I(h) =  I_0(h) \cup I_1(h) \cup I_2(h)$. \index{$I(h)$}

\begin{example}
For example in the figure on the right, $h$ is a hybrid vertex of $\mathcal{H}^a_1$ and the path $P_h$ contains $4$ hybrid vertices $h_1,h_2,h_3,h_4$.
Suppose that $h_2 \in \mathcal{H}^a_0$, $h_1,h_3 \in \mathcal{H}^a_1$,  $h_4 \in \mathcal{H}^a_2$. 
So, $I(h) = \{h_2,h_3,h_4\}$ where  $h_2 \in I_0(h)$, $h_3 \in I_1(h)$, $h_4 \in I_2(h)$. $h_1 \not\in I(h)$ because the path $P_{h_1}$ does not have common vertices above $h_1$ with $P_h$.
\label{ex:I_h}
\end{example}
\end{minipage}\hfill
\begin{minipage}[b]{4.5cm}
\def\svgwidth{4.5cm}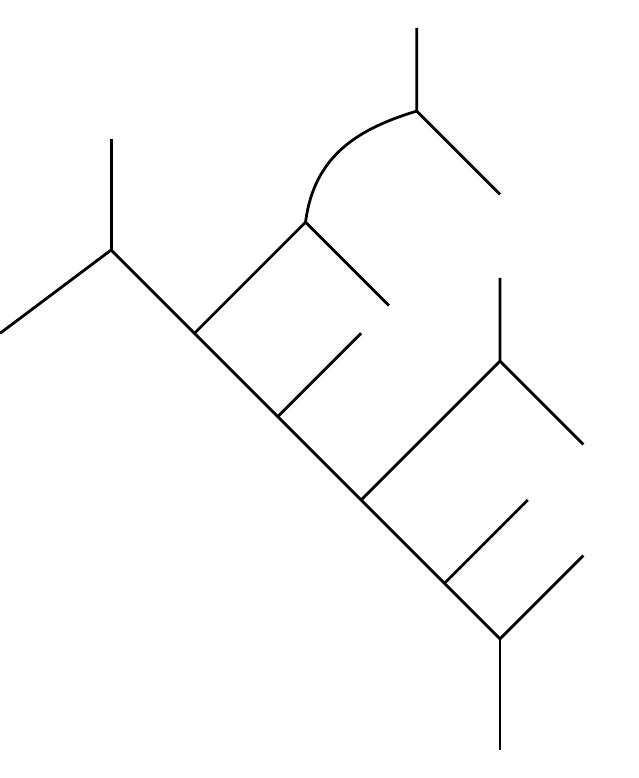
\end{minipage}

Next, the function $\textbf{t}$ is defined as follows.

\begin{definition} \index{function $t$} For every $h \in \mathcal{H}^a_1$, if $I(h) = \emptyset$ then $t(h) = null$. Otherwise, let $h_0$ be the highest hybrid vertex of $I(h)$, so:

If $h_0 \in \mathcal{H}^a_0 \cup \mathcal{H}^a_{\ge 2}$ then $t(h) = h_0$.

If $h_0 \in \mathcal{H}^a_1$, then $t(h) = t(h_0)$, and we denote $h \rightarrow h_0$.
\label{def:t}
\end{definition}

\begin{minipage}[b]{5cm}
\begin{example}
For example, in Figure (i) suppose that $h_1,h_2 \in \mathcal{H}^a_1$ and $h_3 \in \mathcal{H}^a_0$, then $t(h) = h_3$.

In Figure (ii), suppose that $h_1,h_2,h_3 \in \mathcal{H}^a_1$, and $h_0 \in \mathcal{H}^a_0$, then $t(h) = t(h_3)$ and $h \rightarrow h_3$. Next, $t(h_3) = h_0$, so $t(h)=t(h_3) = h_0$.
\end{example}
\end{minipage}\hfill
\begin{minipage}[b]{10cm}
\def\svgwidth{10cm}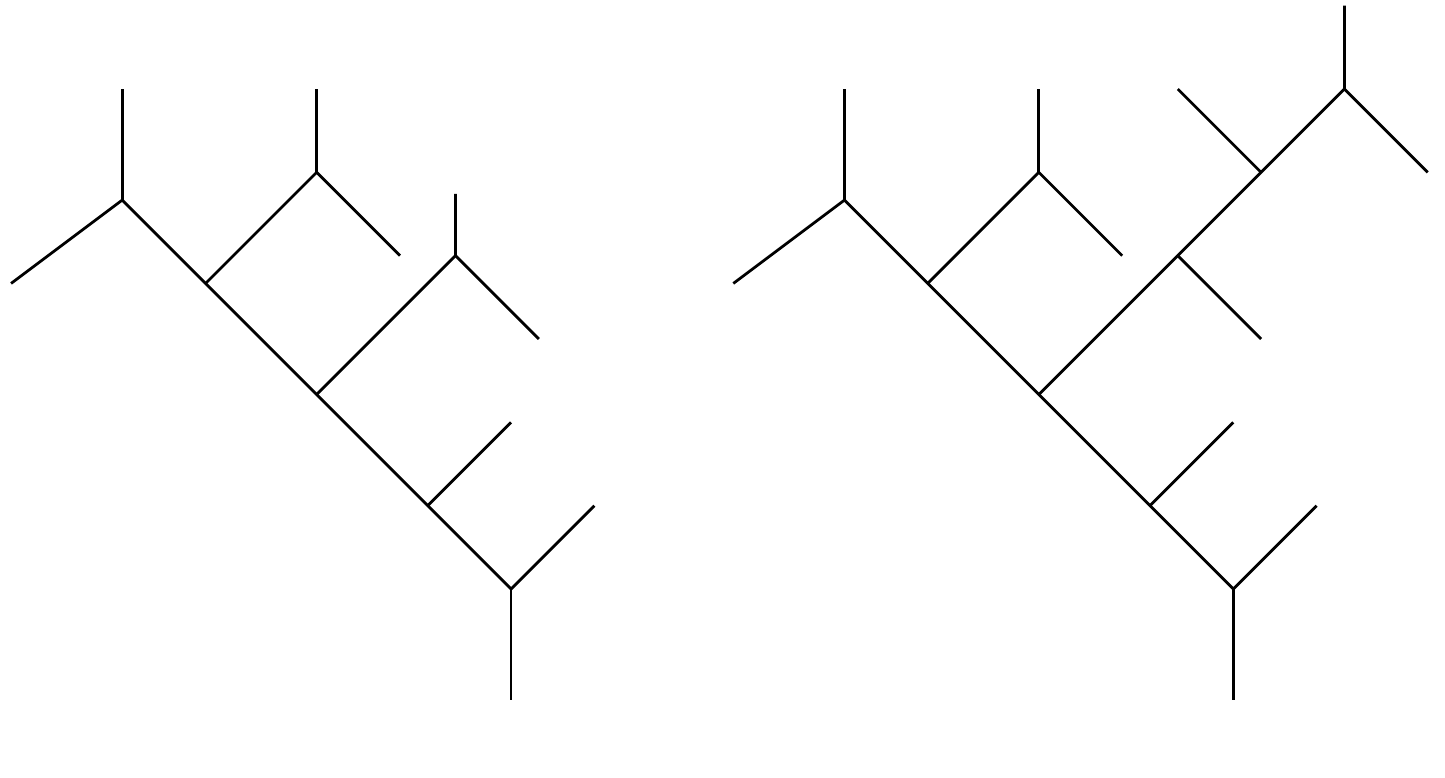
\end{minipage} 

The $3$ following lemmas will be used to prove some properties of the function $t$.

\begin{lemma}
$\forall h \in \mathcal{H}^a_1$, $\forall S \in \mathcal{S}^a_{\le 1}$, if $lca(S)$ is on $P_h$ then there exists a hybrid vertex in $\mathcal{H}^a_0$ which is on $P_h$ and above $lca(S)$.
\label{lem:sup}
\end{lemma}

\begin{pf}
Let $s, s'$ be two children of $S$ such that $lca(s, s') = lca(S)$. Suppose that there is a hybrid vertex on $P_h$ below $lca(S_h)$ and above $lca(S)$, then let $h_0$ be the  lowest one (Figure \ref{fig:Lem_sup_t1}). If $h_0 \in \mathcal{H}^a_0$ then we are done. Suppose that $h_0$ is not in $\mathcal{H}^a_0$, then there exists the split SN-set $S_{h_0} = a^{-1}(h_0)$ and a child $s_0$ of $S_{h_0}$ such that there is a path $h_0 \twoheadrightarrow u_0$. Let $s'_0$ be another child of $S_{h_0}$ such that $lca(s_0,s'_0)$ is above $h_0$. 
Since $h_0$ is the lowest hybrid vertex on $P_h$ below $lca(S_h)$ and above $lca(S)$, then there is no hybrid vertex on $P_h$ which is below $h_0$ and above $lca(S)$. Therefore, in order that $s|s_0s'_0$ is consistent with $N$, there must exist a hybrid vertex below $lca(S)$ and above $u$. Similarly for $s'|s_0s'_0$, there must be a hybrid vertex  below $lca(S)$ and above $u'$. It means that $a(S)$ contains at least two elements, a contradiction. 

\begin{figure}[ht]
\begin{center}
\subfigure[$\exists$ a hybrid vertex $h_0$ not in $\mathcal{H}^a_0$ on $P_h$ above $lca(S)$\label{fig:Lem_sup_t1}]{\def\svgwidth{5cm}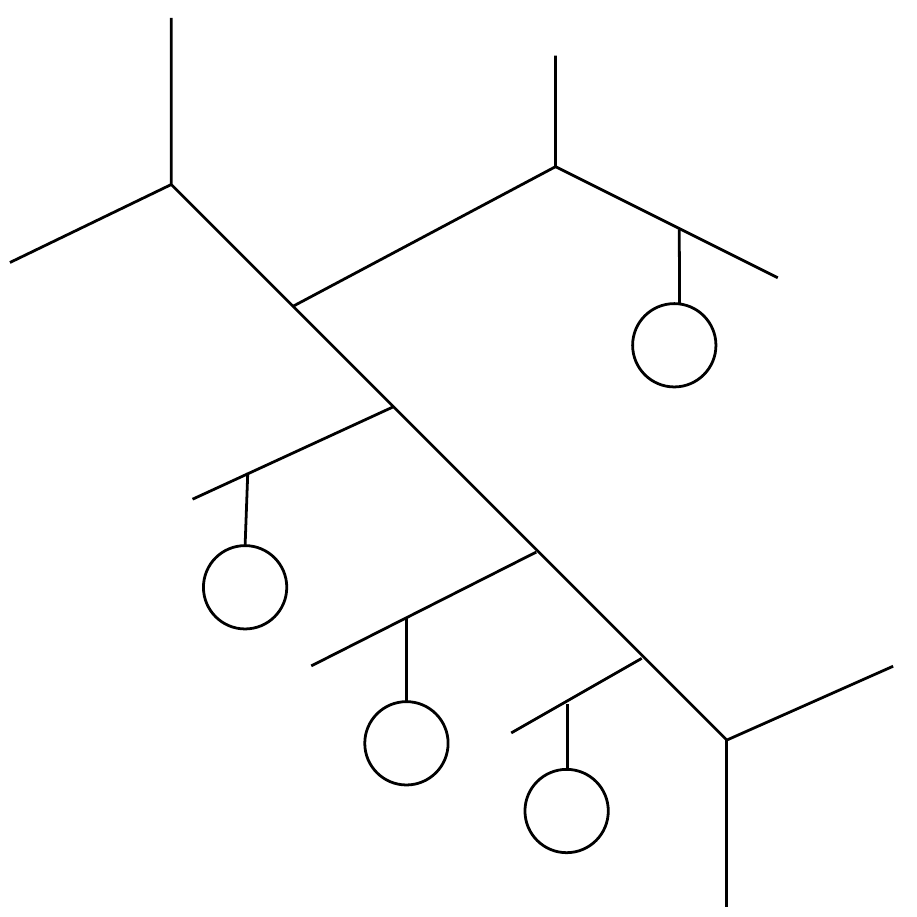}\;\;\;\;\;\;\;
\subfigure[There is not any hybrid vertex on $P_h$ above $lca(S)$\label{fig:Lem_sup_t1_}]{\def\svgwidth{5cm}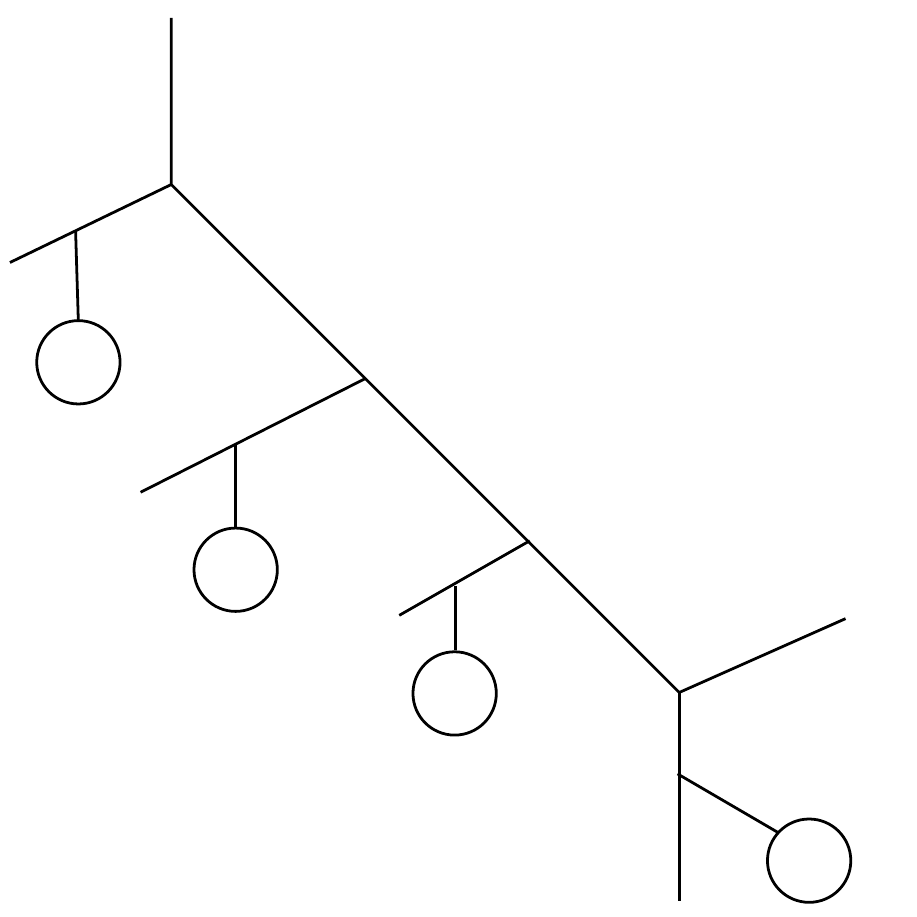}
\caption{Proof of Lemma \ref{lem:sup}}
\label{fig:t_i}
\end{center}
\end{figure}

Hence, there is not any hybrid vertex on $P_h$ below $lca(S_h)$ and above $lca(S)$. Because $h \in a(S_h)$, $S_h$ has a child $s_h$ such that either $h=u_h$ or there is a path $h \twoheadrightarrow u_h$. Let $s'_h$ be another child of $S_h$ such that $lca(s_h,s'_h) = lca(S_h)$ (Figure \ref{fig:Lem_sup_t1_}). So, there is not any hybrid vertex on any path from $lca(S_h)$ to $u'_h$, because otherwise $a(S_h)$ contains another hybrid vertex different from $h$, contradicting $S_h \in \mathcal{S}^a_1$. It implies that every path from a vertex above $lca(S_h)$ to $u'_h$ must pass $lca(S_h)$. Therefore, in order that $s|s_hs'_h$ is consistent with the network, there must exist a hybrid vertex below $lca(S)$ and above $u$. Similarly for $s'|s_hs'_h$,  there must exist a hybrid vertex below $lca(S)$ and above $u'$. It means that $a(S)$ contains at least two elements, a contradiction. 
\end{pf}

\begin{lemma}
For any $h \in \mathcal{S}^a_1$, let $h'$ be a hybrid vertex on $P_h$ which is in $\mathcal{H}^b_2$. Then there is a hybrid vertex of $\mathcal{H}^a_0$ above $h'$ on $P_h$.
\label{lem:sup1}
\end{lemma}

\begin{pf}
Because $h' \in \mathcal{H}^b_2$, there exist $T, T' \in  \mathcal{S}^a_0$ such that $h' \in b(T) \cap b(T')$. 
By definition, $h \not\in N[T]$ and there exist a path $c_T: lca(T) \hookrightarrow h$. Similarly for $T'$ have a path $c_{T'}$.

We will prove that there exists a path $c$ from the root to $lca(S_h)$ which is vertex-disjoint with $N[T]$ (Figure \ref{fig:ap(ii)}). 
Since $h \in a(S_h)$, $S_h$ has a child $s$ such that either $u=h$ or there is path $h \twoheadrightarrow u$. Let $s'$ be another child of $S_h$ such that $lca(s,s')=lca(S_h)$. There is not any hybrid vertex above $u'$ and below $lca(S_h)$ because otherwise $a(S_h)$ contains another hybrid vertex different from $h$. By Lemma \ref{lem:a1} (iii), there is a path $c'$ from the root to $u'$ which is vertex disjoint with $N[T]$. $c'$ must pass $lca(S_h)$ because otherwise there must be a hybrid vertex above $u'$ and below $lca(S)$. Let $c$ be  the subpath of $c'$ from the root to $lca(S_h)$, so $c$ is also vertex-disjoint with $N[T]$.

The $3$ paths $c_T, c_{T'}, P_h$ pass $h'$ while the indegree of $h'$ is $2$, so at least two among them have common vertices above $h'$. If these two paths are $c_T, c_{T'}$, then by the same argument with the proof of Lemma \ref{lem:function_b} where $T,T'$ correspond to $Y,Z$, we deduce a contradiction. So $P_S$ has common vertices with either $c_T$ or $c_{T'}$. Suppose that it is $c_T$, we have the following cases:

\begin{figure}[ht]
\begin{center}
\subfigure[\label{fig:ap(ii)}]{\includegraphics[scale=.8]{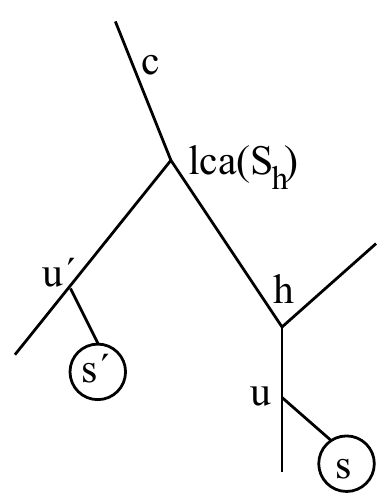}}\;\;\;\;\;
\subfigure[$c_T$ intersects with $P_h$\label{fig:ap(ii)_}]{\includegraphics[scale=.8]{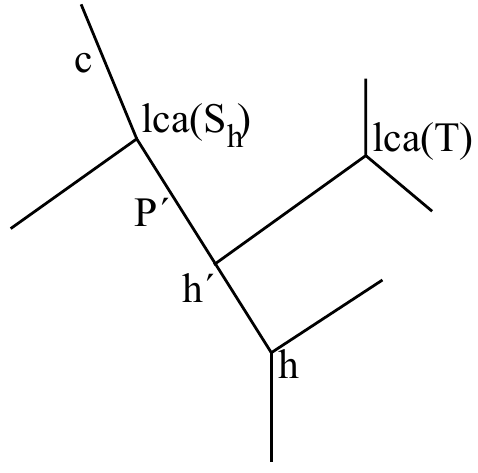}}\;\;\;\;\;
\subfigure[$lca(S_h)$ is on $c_T$\label{fig:ap(ii)1}]{\includegraphics[scale=.8]{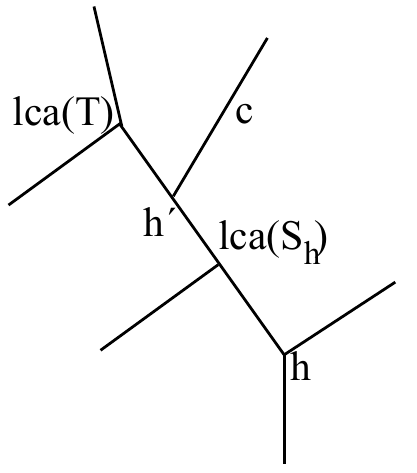}}
\caption{Proof of Lemma \ref{lem:sup1}}
\label{fig:lem_sup}
\end{center}
\end{figure}

- $lca(T)$ is on $P_h$: then  by applying Lemma \ref{lem:sup} with $S=T$, there is a hybrid vertex of $\mathcal{H}^a_0$ on $P_h$ which is above $lca(T)$, so above $h'$, we are done.

- $c_T$ intersects with $P_h$ at a hybrid vertex $h'$ above $h'$ (Figure \ref{fig:ap(ii)_}). As proved above, there is a path $c$ from the root to $lca(S_h)$ which is vertex-disjoint with $N[T]$. Let $P'$ be the subpath of $P_h$ from $lca(S)$ to $h'$, and $C$ be the path consisting of $c$ and $P'$.
We will prove that $lca(T)$ must be on $P'$. 
Because $c_T$ is a path $lca(T) \hookrightarrow h$, every path from the root to $h'$ must have common vertex with $N[T]$. Since $N[T]$ does not contain any hybrid vertex, we deduce from the later that every path from the root to $h'$ must pass $lca(T)$.  In other words, $C$ must pass $lca(T)$. We know that $lca(T)$ can not be on $c$ because this path is vertex disjoint with $N[T]$. So $lca(T)$ must be on $P'$. Hence, we return to the previous case.

- $lca(S_h)$ is on $c_T$ (Figure \ref{fig:ap(ii)1}): As proved above, there is a path $c$ from the root to $lca(S_h)$ which is vertex-disjoint with $N[T]$. This path intersects with $c_T$ at a hybrid vertex $h'$ above $lca(S_h)$. The subpath of $c$ from the root to $h'$ is also vertex disjoint with $N[T]$, contradicting the fact that $c_T$ is a path $lca(T) \hookrightarrow h$. 
\end{pf}

\begin{lemma} 

(i) $\forall h_1,h_2 \in \mathcal{H}^a_1$ if $h_1 \rightarrow h_2$ then $h_2 \in \mathcal{H}^a_1 \cap \mathcal{H}^b_1$.

(ii) $\forall h_1,h_2 \in \mathcal{H}^a_1$ such that $h_1 \neq h_2$, if $h_1 \rightarrow h'_1$ and $h_2 \rightarrow h'_2$ then $h'_1 \neq h'_2$. 

\label{lem:sup2}
\end{lemma}

\begin{pf}

(i) By definition \ref{def:t}, $h_2 \in \mathcal{H}^a_1$, and $h_2$ is the highest hybrid vertex of $I(h_1)$ on $P_{h_1}$. Suppose that $h_2 \in \mathcal{H}^a_1 \cap \mathcal{H}^b_2$, then according to Lemma \ref{lem:sup1}, there is a hybrid vertex of $\mathcal{H}^a_0$ on $P_{h_1}$ above $h_2$. So, this hybrid vertex is in $I_0(h_1)$. It is a contradiction because $h_2$ must be the highest hybrid vertex of $I(h_1)$. Hence, $h_2 \in \mathcal{H}^a_1 \cap \mathcal{H}^b_1$.

\begin{figure}[ht]
\begin{center}
\subfigure[$P_{h_1}$ and $P_{h_2}$ do not have any common vertex above $h_0$\label{fig:t(ii)1}]
{\def\svgwidth{4.2cm}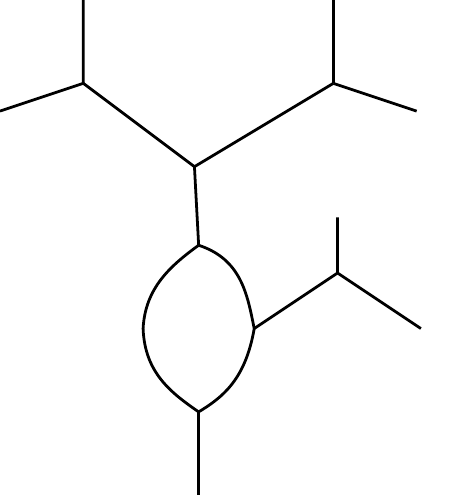}\;\;\;\;
\subfigure[$P_{h_1}$ and $P_{h_2}$ intersect at a hybrid vertex $h'$ above $h_0$\label{fig:t(ii)}]{\def\svgwidth{5.5cm}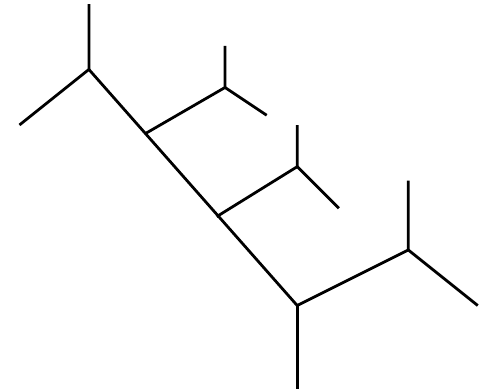}
\caption{Proof of Lemma \ref{lem:sup2} (ii)}
\label{fig:t_ii}
\end{center}
\end{figure}

(ii) Suppose that $h'_1 = h'_2 = h_0$. So, the $2$ paths $P_{h_1}$, $P_{h_2}$ pass $h_0$. There are the following cases:

- $lca(S_{h_2})$ is on $P_{h_1}$, then by Lemma \ref{lem:sup}, there is a hybrid vertex $h'$ of $\mathcal{H}^a_0$ on $P_{h_1}$ above $lca(S_{h_2})$. It means that $h' \in I_0(h_1)$ and is above $h_0$, contradicting $h_0$ being the highest hybrid vertex of $I(h_1)$. Similarly for the case $lca(S_{h_1})$ is on $P_{h_2}$.

- $P_{h_1}$ and $P_{h_2}$ do not have any common vertex above $h_0$ (Figure \ref{fig:t(ii)1}). We have $h_1 \rightarrow h_0$, so $P_{h_1}$ intersects with $P_{h_0}$ at a hybrid vertex $h^1$ above $h_0$. Similarly, $P_{h_2}$ intersects with $P_{h_0}$ at a hybrid vertex $h^2$ above $h_0$. 
Suppose that $h^1$ is above $h^2$, the case where $h^2$ is above $h^1$ will be treated  similarly.
Note that the two paths $P_{h_0}$ and $P_{h_1}$ have two common vertices: $h^1$ and $h_0$. 
Then, by the property that we impose on $P_h$ for any $h$ in $\mathcal{H}^a_1$, the two subpaths of $P_{h_0}$ and $P_{h_1}$ from $h^1$ to $h_0$ are the same. $h^2$ is a vertex on the subpath of $P_{h_0}$ from $h^1$ to $h_0$, so it must be also on the subpath of $P_{h_1}$ from $h^1$ to $h_0$. However, it means that $h^2$ is a common vertex above $h_0$ of $P_{h_1}$ and $P_{h_2}$, a contradiction. 

- $P_{h_1}$ and $P_{h_2}$ intersect at a hybrid vertex $h'$ above $h_0$ (Figure \ref{fig:t(ii)}).  So $h' \not\in I(h_1) \cup I(h_2)$ because $h_0$ is the highest hybrid vertex of $I(h_1)$ and of $I(h_2)$. We deduce that $h' \in \mathcal{H}^a_{\ge 1}$. Let $S'= a^{-1}(h')$.  It is easy to se that $N[S']$ must have common vertices above $h'$ with either $P_{h_1}$ or $P_{h_2}$. So, $h'$ is either in $I(h_1)$ or in $I(h_2)$, a contradiction.

Therefore, $h'_1 \neq h'_2$.
\end{pf}

\begin{lemma} The function $t$ has the following properties:

(i) $\forall h \in \mathcal{H}^a_1 \cap \mathcal{H}^b_2$, $t(h)$ is defined and equal to a hybrid vertex of $(\mathcal{H}^a_0 \cup \mathcal{H}^a_{\ge 2}) \cap \mathcal{H}^b_1$.
 
(ii) $\forall h_0 \in \mathcal{H}^a_0 \cap \mathcal{H}^b_1$, $t^{-1}(h_0)$ contains at most $2$ hybrid vertices of $\mathcal{H}^a_1 \cap \mathcal{H}^b_2$.

(iii) $\forall h_0 \in \mathcal{H}^a_{\ge 2} \cap \mathcal{H}^b_1$, $t^{-1}(h_0)$ contains at most $1$ hybrid vertex of $\mathcal{S}^a_1 \cap \mathcal{H}^b_2$.

\label{lem:function_t}
\end{lemma}

\begin{pf}

(i) Let $h \in \mathcal{H}^a_1 \cap \mathcal{H}^b_2$, suppose that we have a chain of split SN-sets $h \rightarrow h_1 \rightarrow \dots \rightarrow  h_m$ defined as in Definition \ref{def:t}.

Firstly, we will prove that $h_i \neq h$ for any $i$. It is obvious that $h \neq h_1$ because $h \rightarrow h_1$. Suppose that $h_i = h$ for a certain $i>1$, then $h_{i-1} \rightarrow h$. By Lemma \ref{lem:sup2} (i), $h \not\in \mathcal{H}^a_1 \cap \mathcal{H}^b_2$, a contradiction. 

Next, we will prove that $h_i \neq h_j$ for any $i, j = 1, \dots , m$. Suppose otherwise, let $i$ be the smallest index such that there exists $j$ greater than $i$ and $h_i=h_j=h'$. If $i >1$, then we have $h_{i-1} \rightarrow h_i$ and $h_{j-1} \rightarrow h_j$. However, $h_{i-1} \neq h_{j-1}$ because $i$ is the smallest index having this property. So it is a contradiction with Lemma \ref{lem:sup2} (ii). If $i = 1$, we have $h \rightarrow h_i$ and $h_{j-1} \rightarrow h_j$. However, as proved recently, $h \neq h_{j-1}$, so it is a contradiction with Lemma \ref{lem:sup2} (ii). 

Hence, the recursive calls in Definition \ref{def:t} do not loop, and since the number of split SN-Sets is finite, $t(h)$ is always defined.

Now, we show that $\forall h \in \mathcal{H}^a_1 \cap \mathcal{H}^b_2$, $t(h) \neq null$. By Lemma \ref{lem:sup1}, the fact $h \in \mathcal{H}^a_1 \cap \mathcal{H}^b_2$ deduces that $I_0(h) \neq \emptyset$, i.e. $I(S) \neq \emptyset$. Let $h_0$ be the highest vertex of $I(h)$, if $h_0 \in I_0(h) \cup I_2(h)$, then by definition  $t(S) = h_0 \neq null$.
Suppose that $h_0 \in I_1(h)$, and to define $t(h)$ we pass a chain of other split SN-sets: $h \rightarrow h_1 \rightarrow \dots \rightarrow h_i$ and suppose that $I(h_i)=\emptyset$, i.e. $t(h_i) =  null$. Because $h_{i-1} \rightarrow h_i$, the two paths $P_{h_{i-1}}$ and $P_{h_i}$ pass $h_i$ and have common vertices above $h_i$. There are the following cases:

- $lca(S_{h_i})$ is on $P_{h_{i-1}}$. Then by Lemma \ref{lem:sup}, there is a hybrid vertex of $\mathcal{H}^a_0$ on $P_{h_{i-1}}$, above $lca(S_{h_i})$, so above $h_i$. It means that this hybrid vertex is in $I(h_{i-1})$ and above $h_i$, contradicting $h_i$ being the highest hybrid vertex in $I(h_{i-1})$.

- $lca(S_{h_{i-1}})$ is on $P_{h_i}$. Then by Lemma \ref{lem:sup}, there is a hybrid vertex of $\mathcal{H}^a_0$ on $P_{h_i}$ above $lca(S_{h_{i-1}})$. It means that this hybrid vertex is in $I(h_i)$, contradicting $I(h_i) = \emptyset$.

\begin{minipage}[b]{10.5cm}
- $P_{h_i}$ and $P_{h_{i-1}}$ intersect at a hybrid vertex $h'$ above $h_i$ (figure on the right). $h'$ is not in $I(h_{i-1})$ because $h_i$ is the highest vertex of $I(h_{i-1})$. We deduce that $h' \in \mathcal{H}^a_{\ge 1}$. Let $S'= a^{-1}(h')$, so $N[S']$ must have common vertices above $h'$ with either $P_{S_{i-1}}$ or $P_{S_i}$. If it has common vertices with $P_{h_{i-1}}$ then $h'\in I(h_{i-1})$, contradicting $h_i$ being the highest vertex of $I(h_{i-1})$. If it has common vertices with $P_{h_i}$ then $h' \in I(h_i)$, i.e. $I(h_i) \neq \emptyset$, a contradiction.
\end{minipage}\hfill
\begin{minipage}[b]{4cm}
\def\svgwidth{4cm}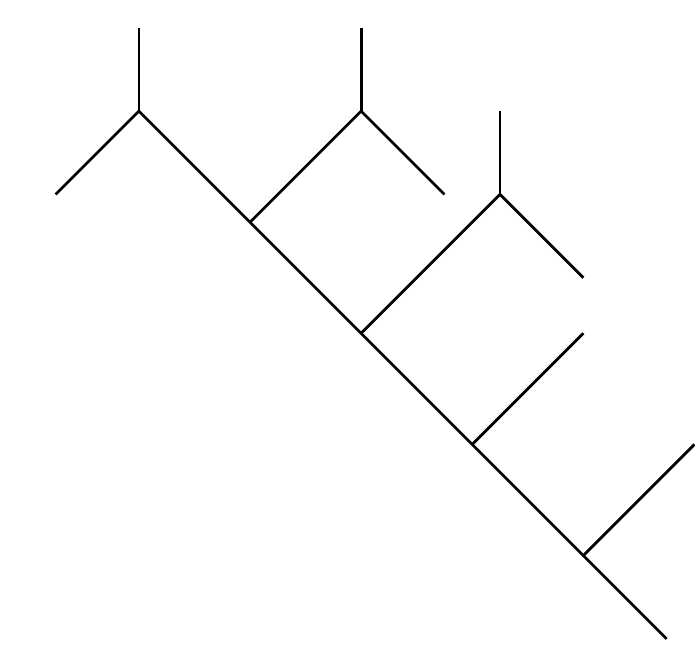
\end{minipage}

So $t(h_i) \neq null$ or $t(h) \neq null$. Let $t(h_i) = h_0$,  we deduce from definition of $t$ that $h_0$ is a hybrid vertex of $\mathcal{H}^a_0 \cup \mathcal{H}^a_{\ge 2}$. If $h_0 \in \mathcal{H}^b_2$, then by Lemma \ref{lem:sup1}, there is a hybrid vertex of $\mathcal{H}^a_0$ on $P_{h_i}$ above $h_0$. Then, this hybrid vertex is in $I_0(h_i)$, contradicting $h_0$ being the highest vertex of $I(h_i)$. Hence, $h_0 \not\in \mathcal{H}^b_2$, i.e. $t(h) \in (\mathcal{H}^a_0 \cup \mathcal{H}^a_{\ge 2}) \cap \mathcal{H}^b_1$.

\begin{figure}[ht]
\begin{minipage}[b]{7cm}
\begin{center}
\def\svgwidth{4cm}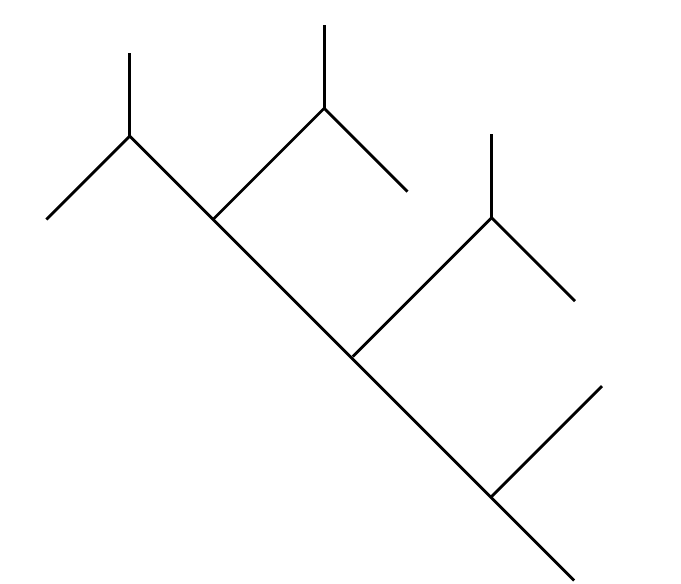
\caption{$h_0 \in \mathcal{H}^a_1\cap \mathcal{H}^b_1$}
\label{fig:t(iii)}
\end{center}
\end{minipage}\hfill
\begin{minipage}[b]{7cm}
\begin{center}
\def\svgwidth{4cm}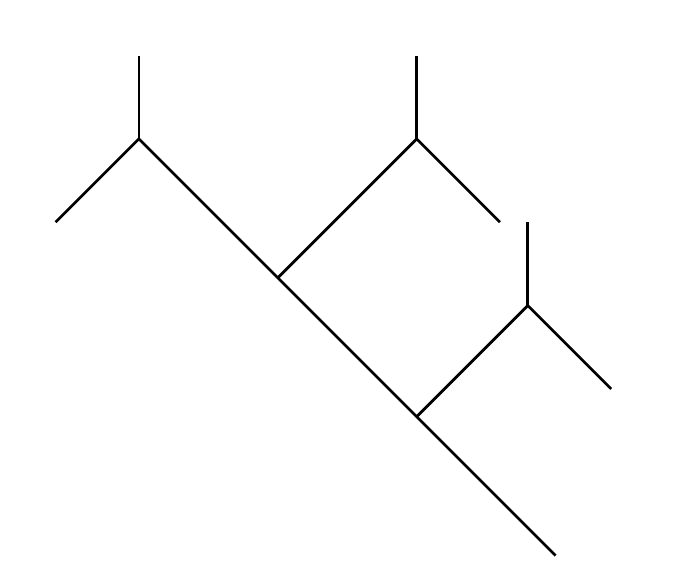
\caption{$h_0 \in \mathcal{H}^a_{\ge 2}\cap \mathcal{H}^b_1$}
\label{fig:t(iv)}
\end{center}
\end{minipage}
\end{figure}

(ii) Let $h_0$ be a hybrid vertex in $\mathcal{H}^a_0 \cap \mathcal{H}^b_1$. Suppose that there are $3$ distinct hybrid vertices $h_1, h_2, h_3$ of $\mathcal{H}^a_1 \cap \mathcal{H}^b_2$ such that $t(h_1) = t(h_2) = t(h_3) = h_0$. 

Suppose that before reaching $h_0$, each one passes a chain of split SN-sets:	

$h_1 \rightarrow \dots \rightarrow h'_1$, $h_2 \rightarrow \dots \rightarrow h'_2$, $h_3 \rightarrow \dots \rightarrow h'_3$, and $t(h'_1) = t(h'_2) = t(h'_3) = h_0$, i.e. $h_0$ is the highest hybrid vertex of $I(h'_1)$, $I(h'_2)$, $I(h'_3)$.

By Lemma \ref{lem:sup2} (i), in each chain, only the first hybrid vertex, i.e. $h_1$, $h_2$, $h_3$, is in $\mathcal{H}^a_1\cap \mathcal{H}^b_2$, the others are not. Because $h_1, h_2, h_3$ are distinct, by Lemma \ref{lem:sup2} (ii) these chains do not have common split SN-sets. In other words, $h'_1$, $h'_2$, $h'_3$ are also distinct. 
So, we need only to show that $h_0$ can not be the highest hybrid vertex of $I(h'_1)$, $I(h'_2)$, $I(h'_3)$. The $3$ paths $P_{h'_1}$, $P_{h'_2}$, $P_{h'_3}$ pass $h_0$, then among them there are at least $2$, for example $P_{h'_1}$ and $P_{h'_2}$, have  common vertices above $h_0$. There are the following cases:

- If $lca(S_{h'_2})$ is on $P_{h'_1}$, then by Lemma \ref{lem:sup}, there is a hybrid vertex of $\mathcal{H}^a_0$ on $P_{h'_1}$, above $lca(S_{h'_2})$, i.e above $h_0$. Then, this hybrid vertex is in $I_{0}(h'_1)$. It is a contradiction because $h_0$ is supposed to be the highest hybrid vertex of $I(h'_1)$. 
Similarly for the case where $lca(S_{h'_1})$ is on $P_{h'_2}$.

- If $P_{h'_1}$ and $P_{h'_2}$ intersect at a hybrid vertex $h'$ above $h_0$ (Figure \ref{fig:t(iii)}), then $h' \not\in I(h'_1) \cup I(h'_2)$. So,  $h' \in \mathcal{H}^a_{\ge 1}$. Let $S'=a^{-1}(h')$, then $N[S']$ must have common vertices above $h'$ with either $P_{h'_1}$, or $P_{h'_2}$. So, either $h' \in I(h'_1)$ or $h \in I(h'_2)$,  a contradiction. 

(iii) Let $h_0$ be a hybrid vertex in $\mathcal{H}^a_{\ge 2} \cap \mathcal{H}^b_1$, and $S_0 = a^{-1}(h_0)$. Suppose that there are $2$ hybrid vertices $h_1, h_2$ of $\mathcal{H}^a_1\cap \mathcal{H}^b_2$ such that $t(h_1) = t(h_2) = h_0$. 

Suppose that before reaching $h_0$, each one passes a chain of split SN-sets: 

$h_1 \rightarrow \dots \rightarrow h'_1$, $h_2 \rightarrow \dots \rightarrow h'_2$, and $t(h'_1) = t(h'_2) = h_0$, i.e. $h_0$ is the highest hybrid vertex of $I(h'_1)$, $I(h'_2)$.
Similarly with Claim (ii), $h_1, h_2$ are the only hybrid vertices of $\mathcal{S}^a_1 \cap \mathcal{H}^b_2$ in these $2$ chains and $h'_1, h'_2$ are distinct. So, we have to only show that  $h_0$ can not be the highest hybrid vertex of $I(h'_1), I(h'_2)$. Suppose otherwise, because $P_{h'_1}$, $P_{h'_2}$ pass $h_0$, we have the following cases.

- If $lca(S_{h'_2})$ is on $P_{h'_1}$, or $lca(S_{h'_1})$ is on $P_{h'_2}$, or $P_{h'_1},P_{h'_2}$ intersect at a hybrid vertex $h'$ above $h_0$, then by the same argument as that of Claim 	(ii), we deduce contradictions.

- The last case is the case where $P_{h'_1}$ and $P_{h'_2}$ intersect at $h_0$ (Figure \ref{fig:t(iv)}). Let $S_0 = a^{-1}(h_0)$, then $N[S_0]$ must have common vertices above $h_0$ with either $P_{h'_1}$ or $P_{h'_2}$. It means that either $h_0 \in I(h'_1)$ or $h_0 \in I(h'_2)$, a contradiction.
\end{pf}

\begin{lemma}
$|\mathcal{H}^b_2| \le 2|\mathcal{H}^a_0|+|\mathcal{H}^a_{\ge 2}|$
\label{lem:f_t}
\end{lemma}

\begin{pf}
By Lemma \ref{lem:function_t}, we deduce that $|\mathcal{H}^a_1\cap \mathcal{H}^b_2|\le 2|\mathcal{H}^a_0\cap \mathcal{H}^b_1|+|\mathcal{H}^a_{\ge 2}\cap \mathcal{H}^b_1|$ (Figure \ref{fig:set}(c))

$\Rightarrow |\mathcal{H}^a_0\cap \mathcal{H}^b_2|+|\mathcal{H}^a_1\cap \mathcal{H}^b_2|+|\mathcal{H}^a_{\ge 2}\cap \mathcal{H}^b_2| \le 
|\mathcal{H}^a_0\cap \mathcal{H}^b_2|+
2|\mathcal{H}^a_0\cap \mathcal{H}^b_1|+|\mathcal{H}^a_{\ge 2}\cap \mathcal{H}^b_1|+|\mathcal{H}^a_{\ge 2}\cap \mathcal{H}^b_2|$

$\Rightarrow |\mathcal{H}^b_2| \le 2|\mathcal{H}^a_0|+|\mathcal{H}^a_{\ge 2}|$

\end{pf}

\subsection{A bound}

\begin{theorem}
\label{theo:bound} If $\mathcal{T}$ is consistent with a level-$k$ network $N$, then there exists a level-$k$ network $N'$ with the same number of hybrid vertices of $N$, which has at most $\lfloor \frac{4}{3}k \rfloor$ split SN-sets.
\end{theorem}

\begin{pf}
According to Lemmas \ref{lem:2} (ii), \ref{lem:f_t}, we have:

$|\mathcal{S}| \le  \frac{4}{3}k  + \frac{1}{3}|\mathcal{H}^b_2| - |\mathcal{H}^a_0| - \frac{1}{2}|\mathcal{H}^a_{\ge 2}| \le \frac{4}{3}k - \frac{1}{6}|\mathcal{H}^b_2| \le \frac{4}{3}k$. 

Therefore, by Lemma \ref{lem:notsplit1} and the assumption stated at the beginning of the section, if $\mathcal{T}$ is consistent with a level-$k$ network $N$, then there exists a level-$k$ network $N'$ with the same number of hybrid vertices of $N$, which has at most $\lfloor \frac{4}{3}k \rfloor$ split SN-sets.
\end{pf}

\textbf{Remark:} 
Up to now we do not have any example achieving this bound. Therefore it is possible that the bound $\lfloor \frac{4}{3}k \rfloor$ is not optimal,  in fact  each time we were able to construct an example of a network which reaches this bound, we can modify it into another network which has a smaller number of split SN-sets without changing the number of hybrid vertices. Especially, for the cases of $k \le 8$, it can be checked case by case that the number of split SN-sets in the restricted networks class is bounded by $k$.

%% file: Figures/function.pdf_tex

\begingroup
  \makeatletter
  \providecommand\color[2][]{%
    \errmessage{(Inkscape) Color is used for the text in Inkscape, but the package 'color.sty' is not loaded}
    \renewcommand\color[2][]{}%
  }
  \providecommand\transparent[1]{%
    \errmessage{(Inkscape) Transparency is used (non-zero) for the text in Inkscape, but the package 'transparent.sty' is not loaded}
    \renewcommand\transparent[1]{}%
  }
  \providecommand\rotatebox[2]{#2}
  \ifx\svgwidth\undefined
    \setlength{\unitlength}{624pt}
  \else
    \setlength{\unitlength}{\svgwidth}
  \fi
  \global\let\svgwidth\undefined
  \makeatother
  \begin{picture}(1,0.41025641)%
    \put(0,0){\includegraphics[width=\unitlength]{Figures/function.pdf}}%
    \put(0.13582683,0.38572253){\color[rgb]{0,0,0}\makebox(0,0)[lb]{\smash{$a$}}}%
    \put(0.4502824,0.08609328){\color[rgb]{0,0,0}\makebox(0,0)[lb]{\smash{$|b| \ge 3$}}}%
    \put(0.21274991,0.29597894){\color[rgb]{0,0,0}\makebox(0,0)[lb]{\smash{$\mathcal{H}^a_0$}}}%
    \put(0.21274991,0.2575174){\color[rgb]{0,0,0}\makebox(0,0)[lb]{\smash{$\mathcal{H}^a_1$}}}%
    \put(0.21274991,0.1549533){\color[rgb]{0,0,0}\makebox(0,0)[lb]{\smash{$\mathcal{H}^a_m$}}}%
    \put(0.12941658,0.27033792){\color[rgb]{0,0,0}\makebox(0,0)[lb]{\smash{$1$}}}%
    \put(0.12941658,0.22546612){\color[rgb]{0,0,0}\makebox(0,0)[lb]{\smash{$i$}}}%
    \put(0.12941658,0.17908441){\color[rgb]{0,0,0}\makebox(0,0)[lb]{\smash{$m$}}}%
    \put(0.52897726,0.30082779){\color[rgb]{0,0,0}\makebox(0,0)[lb]{\smash{$\mathcal{H}^b_1$}}}%
    \put(0.60854934,0.19265542){\color[rgb]{0,0,0}\makebox(0,0)[lb]{\smash{$\mathcal{H}^b_2$}}}%
    \put(0.02044222,0.29597894){\color[rgb]{0,0,0}\makebox(0,0)[lb]{\smash{$\mathcal{S}^a_0$}}}%
    \put(0.02044222,0.2575174){\color[rgb]{0,0,0}\makebox(0,0)[lb]{\smash{$\mathcal{S}^a_1$}}}%
    \put(0.02044222,0.18059433){\color[rgb]{0,0,0}\makebox(0,0)[lb]{\smash{$\mathcal{S}^a_m$}}}%
    \put(0.01329388,0.03988685){\color[rgb]{0,0,0}\makebox(0,0)[lb]{\smash{(a) $|\mathcal{H}^a_i|=i|C_i|$, $\forall i \ge 1$}}}%
    \put(0.33680745,0.03988685){\color[rgb]{0,0,0}\makebox(0,0)[lb]{\smash{(b) $3|\mathcal{S}^a_0| \le |\mathcal{H}^b_1|+2|\mathcal{H}^b_2|$}}}%
    \put(0.68105164,0.04209585){\color[rgb]{0,0,0}\makebox(0,0)[lb]{\smash{(c) $|\mathcal{H}^a_1\cap \mathcal{H}^b_2|$ }}}%
    \put(0.46544342,0.31264561){\color[rgb]{0,0,0}\makebox(0,0)[lb]{\smash{$\le 1$}}}%
    \put(0.46854616,0.24455275){\color[rgb]{0,0,0}\makebox(0,0)[lb]{\smash{$2$}}}%
    \put(0.35762169,0.29213279){\color[rgb]{0,0,0}\makebox(0,0)[lb]{\smash{$\mathcal{S}^a_0$}}}%
    \put(0.4647871,0.39212504){\color[rgb]{0,0,0}\makebox(0,0)[lb]{\smash{$b$}}}%
    \put(0.07848754,0.07968302){\color[rgb]{0,0,0}\makebox(0,0)[lb]{\smash{$|a^{-1}|\le 1$}}}%
    \put(0.84095506,0.39213279){\color[rgb]{0,0,0}\makebox(0,0)[lb]{\smash{$t$}}}%
    \put(0.80890378,0.28654901){\color[rgb]{0,0,0}\makebox(0,0)[lb]{\smash{$\le 2$}}}%
    \put(0.80599044,0.18785281){\color[rgb]{0,0,0}\makebox(0,0)[lb]{\smash{$\le 1$}}}%
    \put(0.69002247,0.25869014){\color[rgb]{0,0,0}\makebox(0,0)[lb]{\smash{$\mathcal{H}^a_1\cap \mathcal{H}^b_2$}}}%
    \put(0.87539321,0.29861902){\color[rgb]{0,0,0}\makebox(0,0)[lb]{\smash{$\mathcal{H}^a_0\cap \mathcal{H}^b_1$}}}%
    \put(0.88167693,0.16601084){\color[rgb]{0,0,0}\makebox(0,0)[lb]{\smash{$\mathcal{H}^a_{\ge 2}\cap \mathcal{H}^b_1$}}}%
    \put(0.78361574,0.08609328){\color[rgb]{0,0,0}\makebox(0,0)[lb]{\smash{$|t|=1$}}}%
    \put(0.68105164,0.01004449){\color[rgb]{0,0,0}\makebox(0,0)[lb]{\smash{$\le 2|\mathcal{H}^a_0\cap \mathcal{H}^b_1|+|\mathcal{H}^a_{\ge 2}\cap \mathcal{H}^b_1|$}}}%
  \end{picture}%
\endgroup

%% file: Figures/I_h.pdf_tex

\begingroup
  \makeatletter
  \providecommand\color[2][]{%
    \errmessage{(Inkscape) Color is used for the text in Inkscape, but the package 'color.sty' is not loaded}
    \renewcommand\color[2][]{}%
  }
  \providecommand\transparent[1]{%
    \errmessage{(Inkscape) Transparency is used (non-zero) for the text in Inkscape, but the package 'transparent.sty' is not loaded}
    \renewcommand\transparent[1]{}%
  }
  \providecommand\rotatebox[2]{#2}
  \ifx\svgwidth\undefined
    \setlength{\unitlength}{184pt}
  \else
    \setlength{\unitlength}{\svgwidth}
  \fi
  \global\let\svgwidth\undefined
  \makeatother
  \begin{picture}(1,1.2173913)%
  \begin{small}  
    \put(0,0){\includegraphics[width=\unitlength]{Figures/I_h.pdf}}%
    \put(0.79318927,0.16909072){\color[rgb]{0,0,0}\makebox(0,0)[lb]{\smash{$h$}}}%
    \put(-0.07685943,0.83623072){\color[rgb]{0,0,0}\makebox(0,0)[lb]{\smash{$lca(S_h)$}}}%
    \put(0.21828079,0.65484246){\color[rgb]{0,0,0}\makebox(0,0)[lb]{\smash{$h_1$}}}%
    \put(0.34125588,0.53801587){\color[rgb]{0,0,0}\makebox(0,0)[lb]{\smash{$h_2$}}}%
    \put(0.47037971,0.39966907){\color[rgb]{0,0,0}\makebox(0,0)[lb]{\smash{$h_3$}}}%
    \put(0.60565232,0.27361953){\color[rgb]{0,0,0}\makebox(0,0)[lb]{\smash{$h_4$}}}%
    \put(0.79841753,0.65832253){\color[rgb]{0,0,0}\makebox(0,0)[lb]{\smash{$lca(S_{h_4})$}}}%
    \put(0.49804913,0.85160256){\color[rgb]{0,0,0}\makebox(0,0)[lb]{\smash{$lca(S_{h_3})$}}}%
    \put(0.6609911,1.05451155){\color[rgb]{0,0,0}\makebox(0,0)[lb]{\smash{$lca(S_{h_1})$}}}%
  \end{small}
  \end{picture}%
\endgroup

%% file: Figures/ex_t.pdf_tex

\begingroup
  \makeatletter
  \providecommand\color[2][]{%
    \errmessage{(Inkscape) Color is used for the text in Inkscape, but the package 'color.sty' is not loaded}
    \renewcommand\color[2][]{}%
  }
  \providecommand\transparent[1]{%
    \errmessage{(Inkscape) Transparency is used (non-zero) for the text in Inkscape, but the package 'transparent.sty' is not loaded}
    \renewcommand\transparent[1]{}%
  }
  \providecommand\rotatebox[2]{#2}
  \ifx\svgwidth\undefined
    \setlength{\unitlength}{416pt}
  \else
    \setlength{\unitlength}{\svgwidth}
  \fi
  \global\let\svgwidth\undefined
  \makeatother
  \begin{picture}(1,0.53846154)%
  \begin{small}
    \put(0,0){\includegraphics[width=\unitlength]{Figures/ex_t.pdf}}%
    \put(0.35850751,0.10932218){\color[rgb]{0,0,0}\makebox(0,0)[lb]{\smash{$h$}}}%
    \put(-0.02632172,0.40440333){\color[rgb]{0,0,0}\makebox(0,0)[lb]{\smash{$lca(S_h)$}}}%
    \put(0.10422107,0.32417391){\color[rgb]{0,0,0}\makebox(0,0)[lb]{\smash{$h_1$}}}%
    \put(0.25612755,0.16818788){\color[rgb]{0,0,0}\makebox(0,0)[lb]{\smash{$h_3$}}}%
    \put(0.1772648,0.24977029){\color[rgb]{0,0,0}\makebox(0,0)[lb]{\smash{$h_2$}}}%
    \put(0.32235847,0.36417471){\color[rgb]{0,0,0}\makebox(0,0)[lb]{\smash{$lca(S_{h_2})$}}}%
    \put(0.22465154,0.42322692){\color[rgb]{0,0,0}\makebox(0,0)[lb]{\smash{$lca(S_{h_1})$}}}%
    \put(0.85850754,0.10932218){\color[rgb]{0,0,0}\makebox(0,0)[lb]{\smash{$h$}}}%
    \put(0.47367827,0.40440333){\color[rgb]{0,0,0}\makebox(0,0)[lb]{\smash{$lca(S_h)$}}}%
    \put(0.60422111,0.32417391){\color[rgb]{0,0,0}\makebox(0,0)[lb]{\smash{$h_1$}}}%
    \put(0.75612758,0.17090735){\color[rgb]{0,0,0}\makebox(0,0)[lb]{\smash{$h_3$}}}%
    \put(0.67726482,0.24977029){\color[rgb]{0,0,0}\makebox(0,0)[lb]{\smash{$h_2$}}}%
    \put(0.82915761,0.35057654){\color[rgb]{0,0,0}\makebox(0,0)[lb]{\smash{$lca(S_{h_2})$}}}%
    \put(0.72465157,0.42322692){\color[rgb]{0,0,0}\makebox(0,0)[lb]{\smash{$lca(S_{h_1})$}}}%
    \put(0.93329773,0.47647376){\color[rgb]{0,0,0}\makebox(0,0)[lb]{\smash{$lca(S_{h_3})$}}}%
    \put(0.8748254,0.39896404){\color[rgb]{0,0,0}\makebox(0,0)[lb]{\smash{$h_0$}}}%
    \put(0.23076923,0.01923076){\color[rgb]{0,0,0}\makebox(0,0)[lb]{\smash{$(i)$}}}%
    \put(0.78846154,0.01923076){\color[rgb]{0,0,0}\makebox(0,0)[lb]{\smash{$(ii)$}}}%
  \end{small}
  \end{picture}%
\endgroup

%% file: Figures/lem_sup.pdf_tex

\begingroup
  \makeatletter
  \providecommand\color[2][]{%
    \errmessage{(Inkscape) Color is used for the text in Inkscape, but the package 'color.sty' is not loaded}
    \renewcommand\color[2][]{}%
  }
  \providecommand\transparent[1]{%
    \errmessage{(Inkscape) Transparency is used (non-zero) for the text in Inkscape, but the package 'transparent.sty' is not loaded}
    \renewcommand\transparent[1]{}%
  }
  \providecommand\rotatebox[2]{#2}
  \ifx\svgwidth\undefined
    \setlength{\unitlength}{264pt}
  \else
    \setlength{\unitlength}{\svgwidth}
  \fi
  \global\let\svgwidth\undefined
  \makeatother
  \begin{picture}(1,1)%
    \put(0,0){\includegraphics[width=\unitlength]{Figures/lem_sup.pdf}}%
    \put(0.79831654,0.13785326){\color[rgb]{0,0,0}\makebox(0,0)[lb]{\smash{$h$}}}%
    \put(-0.05134965,0.80742464){\color[rgb]{0,0,0}\makebox(0,0)[lb]{\smash{$lca(S_h)$}}}%
    \put(0.57112769,0.41565884){\color[rgb]{0,0,0}\makebox(0,0)[lb]{\smash{$lca(S)$}}}%
    \put(0.70295045,0.60418867){\color[rgb]{0,0,0}\makebox(0,0)[lb]{\smash{$s'_0$}}}%
    \put(0.22657082,0.34276067){\color[rgb]{0,0,0}\makebox(0,0)[lb]{\smash{$s_0$}}}%
    \put(0.41537984,0.1684757){\color[rgb]{0,0,0}\makebox(0,0)[lb]{\smash{$s'$}}}%
    \put(0.59256985,0.09585718){\color[rgb]{0,0,0}\makebox(0,0)[lb]{\smash{$s$}}}%
    \put(0.25246254,0.62161735){\color[rgb]{0,0,0}\makebox(0,0)[lb]{\smash{$h_0$}}}%
    \put(0.33114194,0.82204534){\color[rgb]{0,0,0}\makebox(0,0)[lb]{\smash{$lca(S_{h_0})$}}}%
    \put(0.72618847,0.76685494){\color[rgb]{0,0,0}\makebox(0,0)[lb]{\smash{$u'_0$}}}%
    \put(0.42592108,0.56736079){\color[rgb]{0,0,0}\makebox(0,0)[lb]{\smash{$u_0$}}}%
    \put(0.70004568,0.28757064){\color[rgb]{0,0,0}\makebox(0,0)[lb]{\smash{$u$}}}%
    \put(0.42118937,0.33985613){\color[rgb]{0,0,0}\makebox(0,0)[lb]{\smash{$u'$}}}%
  \end{picture}%
\endgroup

%% file: Figures/lem_sup_.pdf_tex

\begingroup
  \makeatletter
  \providecommand\color[2][]{%
    \errmessage{(Inkscape) Color is used for the text in Inkscape, but the package 'color.sty' is not loaded}
    \renewcommand\color[2][]{}%
  }
  \providecommand\transparent[1]{%
    \errmessage{(Inkscape) Transparency is used (non-zero) for the text in Inkscape, but the package 'transparent.sty' is not loaded}
    \renewcommand\transparent[1]{}%
  }
  \providecommand\rotatebox[2]{#2}
  \ifx\svgwidth\undefined
    \setlength{\unitlength}{264pt}
  \else
    \setlength{\unitlength}{\svgwidth}
  \fi
  \global\let\svgwidth\undefined
  \makeatother
  \begin{picture}(1,1)%
    \put(0,0){\includegraphics[width=\unitlength]{Figures/lem_sup_.pdf}}%
    \put(0.7248636,0.26568547){\color[rgb]{0,0,0}\makebox(0,0)[lb]{\smash{$h$}}}%
    \put(0.19931622,0.81017955){\color[rgb]{0,0,0}\makebox(0,0)[lb]{\smash{$lca(S_h)$}}}%
    \put(0.39047505,0.60778086){\color[rgb]{0,0,0}\makebox(0,0)[lb]{\smash{$lca(S)$}}}%
    \put(0.22861855,0.35784264){\color[rgb]{0,0,0}\makebox(0,0)[lb]{\smash{$s'$}}}%
    \put(0.46354125,0.22413792){\color[rgb]{0,0,0}\makebox(0,0)[lb]{\smash{$s$}}}%
    \put(0.46138421,0.37198079){\color[rgb]{0,0,0}\makebox(0,0)[lb]{\smash{$u$}}}%
    \put(0.21990428,0.53242705){\color[rgb]{0,0,0}\makebox(0,0)[lb]{\smash{$u'$}}}%
    \put(0.0417858,0.59021957){\color[rgb]{0,0,0}\makebox(0,0)[lb]{\smash{$s'_h$}}}%
    \put(0.83895957,0.04349882){\color[rgb]{0,0,0}\makebox(0,0)[lb]{\smash{$s_h$}}}%
    \put(0.65707869,0.16858483){\color[rgb]{0,0,0}\makebox(0,0)[lb]{\smash{$u_h$}}}%
    \put(0.00956594,0.77333725){\color[rgb]{0,0,0}\makebox(0,0)[lb]{\smash{$u'_h$}}}%
  \end{picture}%
\endgroup

%% file: Figures/lem_sup3.pdf_tex

\begingroup
  \makeatletter
  \providecommand\color[2][]{%
    \errmessage{(Inkscape) Color is used for the text in Inkscape, but the package 'color.sty' is not loaded}
    \renewcommand\color[2][]{}%
  }
  \providecommand\transparent[1]{%
    \errmessage{(Inkscape) Transparency is used (non-zero) for the text in Inkscape, but the package 'transparent.sty' is not loaded}
    \renewcommand\transparent[1]{}%
  }
  \providecommand\rotatebox[2]{#2}
  \ifx\svgwidth\undefined
    \setlength{\unitlength}{136pt}
  \else
    \setlength{\unitlength}{\svgwidth}
  \fi
  \global\let\svgwidth\undefined
  \makeatother
  \begin{picture}(1,1.05882353)%
    \put(0,0){\includegraphics[width=\unitlength]{Figures/lem_sup3.pdf}}%
    \put(0.42426403,0.12062412){\color[rgb]{0,0,0}\makebox(0,0)[lb]{\smash{$h_0$}}}%
    \put(0.42783235,0.64650913){\color[rgb]{0,0,0}\makebox(0,0)[lb]{\smash{$h^1$}}}%
    \put(0.71899396,0.90023587){\color[rgb]{0,0,0}\makebox(0,0)[lb]{\smash{$lca(S_{h_0})$}}}%
    \put(0.18658409,0.8960765){\color[rgb]{0,0,0}\makebox(0,0)[lb]{\smash{$lca(S_{h_1})$}}}%
    \put(0.71958519,0.49497464){\color[rgb]{0,0,0}\makebox(0,0)[lb]{\smash{$lca(S_{h_2})$}}}%
    \put(0.54488808,0.3077992){\color[rgb]{0,0,0}\makebox(0,0)[lb]{\smash{$h^2$}}}%
  \end{picture}%
\endgroup

%% file: Figures/lem_sup3_.pdf_tex

\begingroup
  \makeatletter
  \providecommand\color[2][]{%
    \errmessage{(Inkscape) Color is used for the text in Inkscape, but the package 'color.sty' is not loaded}
    \renewcommand\color[2][]{}%
  }
  \providecommand\transparent[1]{%
    \errmessage{(Inkscape) Transparency is used (non-zero) for the text in Inkscape, but the package 'transparent.sty' is not loaded}
    \renewcommand\transparent[1]{}%
  }
  \providecommand\rotatebox[2]{#2}
  \ifx\svgwidth\undefined
    \setlength{\unitlength}{144pt}
  \else
    \setlength{\unitlength}{\svgwidth}
  \fi
  \global\let\svgwidth\undefined
  \makeatother
  \begin{picture}(1,0.77777778)%
    \put(0,0){\includegraphics[width=\unitlength]{Figures/lem_sup3_.pdf}}%
    \put(-0.08333333,0.64444443){\color[rgb]{0,0,0}\makebox(0,0)[lb]{\smash{$lca(S_{h_1})$}}}%
    \put(0.45555556,0.6111111){\color[rgb]{0,0,0}\makebox(0,0)[lb]{\smash{$lca(S')$}}}%
    \put(0.60277778,0.45555554){\color[rgb]{0,0,0}\makebox(0,0)[lb]{\smash{$lca(S_{h_2})$}}}%
    \put(0.82222222,0.29166665){\color[rgb]{0,0,0}\makebox(0,0)[lb]{\smash{$lca(S_{h_0})$}}}%
    \put(0.6,0.12222221){\color[rgb]{0,0,0}\makebox(0,0)[lb]{\smash{$h_0$}}}%
    \put(0.36111111,0.29722221){\color[rgb]{0,0,0}\makebox(0,0)[lb]{\smash{$h'$}}}%
  \end{picture}%
\endgroup

%% file: Figures/lem_t.pdf_tex

\begingroup
  \makeatletter
  \providecommand\color[2][]{%
    \errmessage{(Inkscape) Color is used for the text in Inkscape, but the package 'color.sty' is not loaded}
    \renewcommand\color[2][]{}%
  }
  \providecommand\transparent[1]{%
    \errmessage{(Inkscape) Transparency is used (non-zero) for the text in Inkscape, but the package 'transparent.sty' is not loaded}
    \renewcommand\transparent[1]{}%
  }
  \providecommand\rotatebox[2]{#2}
  \ifx\svgwidth\undefined
    \setlength{\unitlength}{200pt}
  \else
    \setlength{\unitlength}{\svgwidth}
  \fi
  \global\let\svgwidth\undefined
  \makeatother
  \begin{picture}(1,0.96)%
    \put(0,0){\includegraphics[width=\unitlength]{Figures/lem_t.pdf}}%
    \put(-0.20910301,0.80399975){\color[rgb]{0,0,0}\makebox(0,0)[lb]{\smash{$lca(S_{h_{i-1}})$}}}%
    \put(0.736,0.68399975){\color[rgb]{0,0,0}\makebox(0,0)[lb]{\smash{$lca(S_{h_i})$}}}%
    \put(0.42744852,0.42399975){\color[rgb]{0,0,0}\makebox(0,0)[lb]{\smash{$h'$}}}%
    \put(0.536,0.81599975){\color[rgb]{0,0,0}\makebox(0,0)[lb]{\smash{$lca(S')$}}}%
    \put(0.59314948,0.28799975){\color[rgb]{0,0,0}\makebox(0,0)[lb]{\smash{$h_i$}}}%
    \put(0.70519598,0.11029882){\color[rgb]{0,0,0}\makebox(0,0)[lb]{\smash{$h_{i-1}$}}}%
  \end{picture}%
\endgroup

%% file: Figures/lem_t_.pdf_tex

\begingroup
  \makeatletter
  \providecommand\color[2][]{%
    \errmessage{(Inkscape) Color is used for the text in Inkscape, but the package 'color.sty' is not loaded}
    \renewcommand\color[2][]{}%
  }
  \providecommand\transparent[1]{%
    \errmessage{(Inkscape) Transparency is used (non-zero) for the text in Inkscape, but the package 'transparent.sty' is not loaded}
    \renewcommand\transparent[1]{}%
  }
  \providecommand\rotatebox[2]{#2}
  \ifx\svgwidth\undefined
    \setlength{\unitlength}{196pt}
  \else
    \setlength{\unitlength}{\svgwidth}
  \fi
  \global\let\svgwidth\undefined
  \makeatother
  \begin{picture}(1,0.85714286)%
    \put(0,0){\includegraphics[width=\unitlength]{Figures/lem_t_.pdf}}%
    \put(-0.15690891,0.66471968){\color[rgb]{0,0,0}\makebox(0,0)[lb]{\smash{$lca(S_{h'_1})$}}}%
    \put(0.73870439,0.54114516){\color[rgb]{0,0,0}\makebox(0,0)[lb]{\smash{$lca(S_{h'_2})$}}}%
    \put(0.42385593,0.28282943){\color[rgb]{0,0,0}\makebox(0,0)[lb]{\smash{$h'$}}}%
    \put(0.49268089,0.7142857){\color[rgb]{0,0,0}\makebox(0,0)[lb]{\smash{$lca(S')$}}}%
    \put(0.6151536,0.07931032){\color[rgb]{0,0,0}\makebox(0,0)[lb]{\smash{$h_0$}}}%
  \end{picture}%
\endgroup

%% file: Figures/lem_t__.pdf_tex

\begingroup
  \makeatletter
  \providecommand\color[2][]{%
    \errmessage{(Inkscape) Color is used for the text in Inkscape, but the package 'color.sty' is not loaded}
    \renewcommand\color[2][]{}%
  }
  \providecommand\transparent[1]{%
    \errmessage{(Inkscape) Transparency is used (non-zero) for the text in Inkscape, but the package 'transparent.sty' is not loaded}
    \renewcommand\transparent[1]{}%
  }
  \providecommand\rotatebox[2]{#2}
  \ifx\svgwidth\undefined
    \setlength{\unitlength}{196pt}
  \else
    \setlength{\unitlength}{\svgwidth}
  \fi
  \global\let\svgwidth\undefined
  \makeatother
  \begin{picture}(1,0.85714286)%
    \put(0,0){\includegraphics[width=\unitlength]{Figures/lem_t__.pdf}}%
    \put(-0.14991861,0.65423409){\color[rgb]{0,0,0}\makebox(0,0)[lb]{\smash{$lca(S_{h'_1})$}}}%
    \put(0.78836527,0.41165845){\color[rgb]{0,0,0}\makebox(0,0)[lb]{\smash{$lca(S_{h'_2})$}}}%
    \put(0.51663168,0.18719182){\color[rgb]{0,0,0}\makebox(0,0)[lb]{\smash{$h_0$}}}%
    \put(0.63025332,0.65945869){\color[rgb]{0,0,0}\makebox(0,0)[lb]{\smash{$lca(S_0)$}}}%
  \end{picture}%
\endgroup

%% file: Construction.tex
\section{Constructing a minimum phylogenetic network.}

\restylealgo{boxed}\linesnumbered
\begin{algorithm}[h]
\caption{Constructing a minimum level-$k$ phylogenetic network}
\label{algorithm1}
\KwData{A dense triplet set $\mathcal{T}$ on the set $\mathcal{L}$ of $n$ species, and a fixed $k$}
\KwResult{A minimum level-$k$ network consistent with $\mathcal{T}$, if there exists one}
Compute the SN-tree of $\mathcal{T}$\;
For every singleton $u$ of $\mathcal{L}$, let $N^m_u$ be the network containing only one leaf $u$\;
\For {(each non-singleton SN-set $A$ of $\mathcal{T}$, in non-decreasing order of size)}{
	$\mathcal{T'} = \mathcal{T}|A$;$N^m_A = null$;$min=0$\;
	\For {(each set $\mathcal{C}$ of at most $\lfloor\frac{4}{3}k \rfloor$ disjoint non-singleton descendants of $A$)}{
			$\mathcal{P} \leftarrow$ the partition of $A$ inferred from $\mathcal{C}$\;		
			$N^m_A \leftarrow$ a level-$k$ network consistent with $\mathcal{T}$ and has $\mathcal{P}$ as its partition\;
			$min \leftarrow$ the number of hybrid vertices in $N^m_A$\;
			\For {(each level-$k$ network $N_A$ consistent with $\mathcal{T}$ and has $\mathcal{P}$ as its partition)}{				
				\If{(the number of hybrid vertices of $N_A < min$ )}{
					$min \leftarrow$ the number of hybrid vertices of $N_A$\;
					$N^m_A = N_A$\;
				}
			}					
	}	
	\If {($N^m_A=null$)}{
		\Return null\;
	}
}
\textbf{return} $N^m_{\mathcal{L}}$
\end{algorithm}

\begin{theorem}
For every $\mathcal{T}$ set of dense triplets and a fixed $k$, algorithm \ref{algorithm1} takes time $O(|\mathcal{T}|^{k+1}n^{\lfloor \frac{4k}{3}\rfloor +1})$ and return a minimum level-$k$ network consistent with $\mathcal{T}$ if there is any.
\end{theorem}

\begin{pf}
\subsubsection*{The correctness of Algorithm \ref{algorithm1}}

This algorithm consists of constructing a network on each SN-set following a non-decreasing order of size (the loop For at line $3$). So for each iteration corresponding to a SN-set $A$, a minimum level-$k$ network on each SN-set smaller than $A$ is already constructed. 
Remark that if $N$ is a minimum level-$k$ network consistent with $\mathcal{T}$ then each $N_i$ is a minimum level-$k$ network consistent with $\mathcal{T}|l(N_i)$. So, by constructing for each $A$ a minimum level-$k$ network $N^m_A$ consistent with $\mathcal{T}|A$, finally $N^m_{\mathcal{L}}$ is a minimum level-$k$ network consistent with $\mathcal{T}$.

For each SN-set $A$, we must find a partition of $A$ which is the one in a minimum level-$k$ network consistent with $\mathcal{T}|A$. By Lemma \ref{lem:calculation}, each partition is determined by a set of split SN-sets, and each one is a non-singleton descendant of $A$. By Theorem \ref{theo:bound},  we need only to check all the possible sets of descendants of $A$ having cardinality at most $\lfloor\frac{4}{3}k \rfloor$. That is what the loop For at line $5$ does. Next, for each partition $\mathcal{P}$ inferred from each set of split SN-sets, the algorithm checks all level-$k$ networks which are consistent with $\mathcal{T}$ and have $\mathcal{T}$ as their partition, and then chooses the one which contains the minimum number of hybrid vertices. The finding network is stocked in $N^m_A$. That is what the loop For at line $9$ does. If $N^m_A = null$, i.e there is not any level-$k$ network consistent with $\mathcal{T}|A$, then we conclude that there is not any level-$k$ network consistent with $\mathcal{T}$. 

\subsubsection*{The complexity of Algorithm \ref{algorithm1}}

\hspace*{.6cm}- The SN-tree of $\mathcal{T}$ can be computed  in $O(n^3)$ (using the algorithm in \cite{JNS06}).

- The first loop For: There are at most $O(n)$ non-singleton SN-sets, so there are at most $O(n)$ constructions.

- The second loop For repeats at most $n^{\lfloor\frac{4}{3}k\rfloor}$ times because $A$ has at most $O(n)$ non-singletons descendants, so there are at most $n^{\lfloor\frac{4}{3}k\rfloor}$ possibilities for $\mathcal{C}$.

- In the body of the second loop For: based on Lemma \ref{lem:decomposition}, to construct a level-$k$ network consistent with $\mathcal{T}'$ and has $\mathcal{P}$ as its partition, there are two steps: First, we compute a level-$k$ simple network $N_S$ consistent with $\mathcal{T'} \nabla \mathcal{P}$. Then, we replace each leaf of $N_S$ by the  subnetwork already found on the corresponding part of $\mathcal{P}$. According to \cite{IKM08}, we can compute all level-$k$ simple networks consistent with 
$\mathcal{T'} \nabla \mathcal{P}$ in time $O(|\mathcal{T'} \nabla \mathcal{P}|^{k+1})$, or in time $O(|\mathcal{T}|^{k+1})$. The times needed to compute the partition $\mathcal{P}$ of $A$ from the set of split SN-sets $\mathcal{C}$ (Lemma \ref{lem:calculation}), to replace each leaf of $N_S$ by a subnetwork, are negligible compared to the time for computing all the simple networks. So this loop takes time $O(|\mathcal{T}|^{k+1})$.

Hence, the total complexity is $O(|\mathcal{T}|^{k+1}n^{\lfloor \frac{4k}{3}\rfloor +1})$.
\end{pf}

\begin{corollary}For every $\mathcal{T}$ set of dense triplets, it is polynomial  to compute  a minimum level phylogenetic network consistent with 
$\mathcal{T}$ which minimizes the number of hybrid vertices if the minimum level is restricted.
\end{corollary}

\begin{pf}
It is easy to see that we can slightly modify Algorithm \ref{algorithm1} so that it returns a minimum level network. Indeed, we try to construct a minimum level-$i$ network consistent with $\mathcal{T}$ if there is any, in increasing order of $i$. Then, the first value of $i$ that the algorithm returns a network  corresponds to the minimum level of the networks consistent with $\mathcal{T}$. So the complexity is $O(k|\mathcal{T}|^{k+1}n^{\lfloor \frac{4}{3}k \rfloor +1})$ where $k$ is the minimum level.
\end{pf}

%% file: Conclusion.tex
\section{Conclusion and perspectives}
We proved that for any $k$ fixed, we can construct a level-$k$ network having the minimum number of hybrid vertices,  if there exists one, in polynomial time. Furthermore, if the minimum level of the networks consistent with $\mathcal{T}$ is restricted, we can also construct one in polynomial time.

\cite{IKKS08}  implemented the algorithm for level-$2$ networks and applied it to some part of yeast genomic data. However, on a bigger data set, there does not exist any level-$2$ networks. So with our result, one could expect to practically find solution on real data, for small values of $k$ (as for example with $k \leq 5$).  

For simple networks,  \cite{JS06,JNS06, IKKS08} showed, there are efficient algorithms for level-$1$, level-$2$. However, for general level-$k$ networks, there exists only a $O(|\mathcal{T}|^{k+1})$ algorithm \cite{IKM08}. Any improvement  for this problem, even on small levels, will allow us to implement more efficiently our algorithm.
 
For any triplet set $\mathcal{T}$ we can define its $Treerank(\mathcal{T})$ as the minimum $k$ for which there exists a level-$k$ network representing $\mathcal{T}$. This measures the distance from $\mathcal{T}$ to be consistent with  a tree. This distance is measured  in terms of the number of hybrid vertices.
We proved in this paper that for dense triplet sets, and for any fixed $k$, checking if $Treerank(\mathcal{T})\leq k$ can be done in polynomial time. Furthermore \cite{IK11}
proved the NP-hardness of the computation of the $Treerank(\mathcal{T})\leq k$. Therefore this parameter has a similar behavior on phylogenetic networks that  treewidth or  undirected graphs.  Perhaps this analogy could yield further interesting structural insights as shown in \cite{GBP09} with a nice recursive construction.

Another question is under which conditions on the triplet set $\mathcal{T}$ there is only one network consistent with $\mathcal{T}$. It would be also interesting to know whether the condition of density on the triplet set can be relaxed  so that there is still a polynomial algorithm to construct a consistent level-$k$ network, if there is any, with any fixed $k$.